\documentclass{elsarticle}               

\usepackage{amsmath,amssymb,epsfig}
\usepackage{graphics}
\usepackage{changebar}
\usepackage{graphicx}
\usepackage{graphics}
\usepackage{subfigure}
\usepackage[dvips]{color}
\usepackage{bm}

\journal{Journal of Computational Physics}
\psfigdriver{dvips}

\def\Secref#1{Section~\ref{#1}}

\def\secref#1{section~\ref{#1}}

\def\fig#1{Figure~\ref{#1}}
 
\def\figref#1{figure~\ref{#1}}

\def\tabref#1{Table~\ref{#1}}

\def\eqref#1{Eq.~(\ref{#1})}
\def\Eq.#1{Eq.~(\ref{#1})}
\def\Eqs.#1#2{Eqs.~(\ref{#1}--\ref{#2})} 
\def\Eq#1{Equation~(\ref{#1})}
\def\Eqs#1#2{Equations~(\ref{#1}--\ref{#2})} 
\def\eq#1{equation~(\ref{#1})}
\def\eqs#1#2{equations~(\ref{#1}--\ref{#2})}

\def\bea{\begin{eqnarray}}
\def\eea{\end{eqnarray}}

\def\be{\begin{equation}}
\def\ee{\end{equation}}

\def\ez{\bm{e}_{z}}
\def\bhat{\hat{\bm b}}

\def\Lperp{L_\perp}
\def\kperp{k_\perp}
\def\kpar{k_\parallel}

\def\lapperp{\nabla^2_\perp}
\def\Lpar{L_\parallel}

\def\rhoi{\rho_{i}}
\def\rhoe{\rho_{e}}
\def\rhos{\rho_s}

\def\({\left(}
\def\){\right)}
\def\[{\left[}
\def\]{\right]}
\def\<{\left\langle}
\def\>{\right\rangle}

\def\GG{\mathcal G}
\def\AA{\mathcal A}

\def\NN{\mathcal N}
\def\deta{\mathcal D_\eta}

\def\d{\partial}
\def\D'{\Delta'}
\def\Apar{A_{\parallel}}
\def\ue{u_{e\parallel}}

\def\etal{\textit{et al.}}

\def\dt{\Delta t}

\def\tA{\tau_A}

\def\noe{n_{0e}}

\def\viriato{{\tt Viriato}}
\def\kmax{k_{\rm max}}

\newcommand{\vpar}{v_{\parallel}}
\newcommand{\vthe}{v_{{\rm th}e}}
\newcommand{\Tpar}{T_{\parallel e}}

\newcommand{\lt}{\left}
\newcommand{\rt}{\right}
\newcommand{\vthi}{v_{{\rm th}i}}
\newcommand{\od}[2]{\ensuremath{\frac{d #1}{d #2}}}

\begin{document}
\begin{frontmatter}
  

  \title{\viriato: a Fourier-Hermite spectral code for strongly magnetised fluid-kinetic plasma
dynamics}

  \author[IPFN]{N.~F.~Loureiro\corref{cor}}
  \ead{nloureiro@ipfn.ist.utl.pt}
  \ead[url]{http://web.ist.utl.pt/nuno.f.loureiro}
  \cortext[cor]{Corresponding author}
  \author[UMD]{W.~Dorland}
  \author[IPFN]{L.~Fazendeiro}
  \author[UMD]{A.~Kanekar}
  \author[Oxford]{A.~Mallet}
  \author[IPFN]{M.~S.~Vilelas}
  \author[IPP]{A.~Zocco}
  \address[IPFN]{Instituto de Plasmas e 
    Fus\~ao Nuclear, Instituto Superior T\'ecnico, Universidade de Lisboa,
    1049-001 Lisboa, Portugal}
  \address[UMD]{IREAP \& Department of Physics, University of Maryland, College Park, 
MD 20742, USA}
  \address[Oxford]{Rudolf Peierls Centre for Theoretical Physics, 
University of Oxford, 
    Oxford OX1 3NP, United Kingdom}
    \address[IPP]{Max-Planck-Institut f\"ur Plasmaphysik, Wendelsteinstrasse, D-17489, Greifswald, Germany}

%

\begin{abstract}
We report on the algorithms and numerical methods used in \viriato,
a novel fluid-kinetic code that solves two distinct sets of equations: 
(i) the Kinetic Reduced Electron
Heating Model (KREHM) equations [Zocco \& Schekochihin, 
Phys. Plasmas {\bf 18}, 102309 (2011)] 
(which reduce to the standard Reduced-MHD equations in the appropriate limit)
and (ii) the kinetic reduced MHD (KRMHD) equations 
[Schekochihin \etal, Astrophys. J. Suppl. {\bf 182}:310 (2009)]. 
Two main applications of these equations are magnetised (Alfv\'enic) plasma 
turbulence and magnetic reconnection.
\viriato~uses operator splitting (Strang or Godunov) to separate the dynamics 
parallel and perpendicular to the
ambient magnetic field (assumed strong). 
Along the magnetic field, \viriato~allows for either a second-order accurate 
MacCormack method or, for higher accuracy, a spectral-like scheme composed of the 
combination of a total variation diminishing (TVD) third order Runge-Kutta method 
for the time derivative with a 7th order upwind scheme for the fluxes. 
Perpendicular to the field \viriato~is pseudo-spectral, 
and the time integration is performed by means of an iterative predictor-corrector 
scheme.
In addition, a distinctive feature of \viriato~is its spectral representation of the 
parallel velocity-space dependence, achieved by means of a Hermite 
representation of the perturbed distribution function. 
A series of linear and nonlinear benchmarks and tests are presented, including a detailed analysis of 2D and 3D Orszag-Tang-type decaying turbulence, both in fluid and kinetic regimes.

\end{abstract}

\end{frontmatter}

\section {Introduction}
\label{sec:intro}
Magnetised plasma dynamics lies at the heart of many fascinating phenomena in astro, space and laboratory physics. 
Turbulence in the solar wind~\cite{bruno_solar_2005} and in the interstellar medium~\cite{elmegreen_interslar_2004}, 
solar~\cite{shibata_solar_2011}, stellar~\cite{haisch_flares_1991} and accretion disk flares~\cite{uzdensky_magnetic_2004}, substorms in the Earth's magnetosphere~\cite{schindler_theory_1974}, and turbulent transport and instabilities in
magnetised fusion experiments~\cite{wesson_tokamaks_2011},
are just a few examples of remarkable physics problems whose solution
is indeed determined by understanding the behaviour of plasmas in a magnetised environment.

In many of these cases, (i) the collision frequency is much lower than the
typical frequencies of the physical phenomena of interest (e.g., turbulence, 
magnetic reconnection) 
--- i.e., the plasmas are weakly collisional; and
(ii) the size of the ion Larmor orbit is several orders of magnitude smaller than the size of the system.
Weak collisionality implies that on the timescales of interest the plasma cannot be
treated as a fluid, and instead a kinetic description that evolves
the particles' distribution functions is
required. 
This is rather unfortunate from the computational point of view, since fully kinetic models
live on a six-dimensional phase-space (each particle is characterised by its
position and velocity vectors). 
The strong magnetisation, however, implies that the plasma is highly
anisotropic, with very different particle motions along and across the magnetic field direction. 
This anisotropy can be explored analytically to yield reduced kinetic models, i.e., 
asymptotic descriptions that
reduce the phase-space to only 5D or even 4D.
This leads to tremendous computational savings and effectively
renders possible calculations that would otherwise not be feasible on today's supercomputers.

Gyrokinetics~\cite{frieman_nonlinear_1982,howes_astrophysical_2006,
garbet_gyrokinetic_2010,
krommes_gyrokinetic_2012, abel_multiscale_2012} 
is a rigorous description of strongly magnetised, weakly-collisional plasmas.
The key idea behind the gyrokinetic formalism is that, because of the strong 
magnetic (guide) field,
the particles' Larmor gyration frequency is much higher than the 
frequencies of dynamical interest, and can thus be averaged over. 
This allows for the reduction of the dimensionality of the system, from 6D 
(three position and three velocity coordinates)
to 5D (three position coordinates, and velocities parallel and perpendicular to the 
magnetic field) while retaining all the essential physical effects.
Gyrokinetics was originally 
motivated by the attempt to model microinstabilities in magnetic fusion 
experiments; in this respect it has been rather successful~\cite{garbet_gyrokinetic_2010,krommes_gyrokinetic_2012}. 
As recognition of its usefulness, the range of applications of gyrokinetics has 
broadened in recent years; it is now routinely applied to the study of 
turbulence in magnetised astrophysical 
systems~\cite{howes_astrophysical_2006,howes_kinetic_2008,
schekochihin_astrophysical_2009,howes_gyrokinetic_2011},
and there have also been some studies pioneering its application to 
the problem of magnetic reconnection~\cite{rogers_gyrokinetic_2007, 
pueschel_gyrokinetic_2011, numata_gyrokinetic_2011, tenbarge_collisionless_2014,
munoz_gk_2015}.

This reduction of the dimensionality of the
system allowed by gyrokinetics is extremely advantageous from the numerical point of view.
Nonetheless, intrinsically multiscale problems such as kinetic turbulence and reconnection 
remain formidable computational challenges.
Further simplification where possible is therefore desirable. 

One possible such simplification of gyrokinetics has recently been proposed by Zocco and 
Schekochihin~\cite{zocco_reduced_2011}: the Kinetic Reduced Electron Heating Model (KREHM), a rigorous asymptotic limit of gyrokinetics valid for plasmas such that
\be
\label{KREHM_order}
\beta_e\sim m_e/m_i,
\ee
where $\beta_e=8\pi n_{0e}T_{0e}/B_0^2$ is the electron beta, $n_{0e},~T_{0e}$ are the 
background electron density and temperature, respectively, and $B_0$ is the background 
magnetic field. 
Under this assumption, Ref.~\cite{zocco_reduced_2011} shows that it is possible to reduce the plasma dynamics 
to a 4D phase-space --- position and velocity parallel to the magnetic field --- while 
retaining key physics such as phase-mixing and electron Landau damping, ion 
finite Larmor radius effects, electron inertia, electron collisions and Ohmic resistivity.
This is a very significant simplification of the full kinetic description, 
which renders possible truly multiscale kinetic simulations. 
In particular, because no {\it ad hoc} fluid closure is employed, KREHM
can be used for detailed studies of energy conversion and dissipation in kinetic  turbulence and reconnection, including electron heating via phase-mixing and Landau damping.

If taken literally, the ordering imposed by \eq{KREHM_order} is somewhat restrictive, and 
obviously excludes many plasmas of interest. Examples of plasmas where it may hold 
are some regions of the solar 
corona~\cite{aschwanden_new_2001,uzdensky_fast_2007}, the LArge Plasma Device (LAPD) experiment at 
UCLA~\cite{gekelman_design_1991}, and edge regions in some tokamaks~\cite{saibene_h-mode_2007}\footnote{We hasten to add that it is unclear whether the fundamental approximations of standard gyrokinetics are at all valid in the edge region of tokamaks; however, if they are, then KREHM may be a good approximation there given that $\beta_e$ does tend to be rather low in this region.}.
However, one may legitimately expect that the plasma behaviour captured by 
KREHM will qualitatively hold beyond its rigorous limits of applicability, 
as is so often 
the case with many other simplified plasma models 
(MHD being a notorious example of a description 
known to work rather well far outside its strict limits of validity). 
Hints that this may indeed be the case are offered in~\secref{sec:tearing}, 
where a direct comparison of KREHM with a (non-reduced) gyrokinetic model for the linear 
collisionless tearing mode problem yields very good agreement at values of $\beta_e$
significantly larger than $m_e/m_i$.

This paper reports on the numerical methods and algorithms used in  
\viriato, the first numerical code to solve this particular set of equations. 
An extensive series of tests and benchmarks is also presented.
Considerable attention is devoted to Orszag-Tang-type decaying turbulence, both in the fluid and kinetic regimes.
The reader interested in the application of this code and physics model 
to the problem of magnetic reconnection is 
referred to~\cite{loureiro_fast_2013}, where the importance of electron heating 
via Landau damping
in reconnection is demonstrated for the first time.

A second set of equations solved by \viriato~are the kinetic reduced MHD (KRMHD) equations~\cite{schekochihin_astrophysical_2009}, which describe the evolution of 
compressible fluctuations 
(density and parallel magnetic field) in the regime $\kperp\rhoi\ll1$ ($\kperp$ being 
the wave number perpendicular to the guide-field of 
a typical perturbation, and $\rhoi$ the ion Larmor radius.)
These equations are structurally identical to those of KREHM, so their numerical implementation 
in \viriato~is straightforward. We also note that KREHM reduces to the 
standard Reduced-MHD (RMHD) set of equations~\cite{kadomtsev_nonlinear_1974,strauss_nonlinear_1976} in the appropriate 
limit (i.e., when the wave length of the fluctuations is much larger than all the 
kinetic scales). Thus, \viriato~can also be used as a RMHD code (in either 2D or 3D slab geometry). 
Finally, we remark that under an isothermal closure for the electrons, KREHM reduces to
the simple two-field gyrofluid model treated in Refs.~\cite{loureiro_studies_2005,
loureiro_iterative_2008} (which 
is a limit of the more complete models of 
Snyder \etal~\cite{snyder_landau_1997} 
and of Schep \etal~\cite{schep_generalized_1994}).

This paper is organized as follows. \Secref{sec:eqs} presents the different sets of equations 
integrated by \viriato. The kinetic equations are solved by means of a Hermite expansion, which requires some form of closure (or truncation). This is discussed in \secref{sec:closure}, where an
asymptotically exact nonlinear closure for the Hermite-moment hierarchy is derived.
\Secref{sec:energy} presents the energy evolution equation for the closed KREHM model; and the normalizations that we adopt are laid out in \secref{sec:KREHM_norms}.
\Secref{sec:numerics} deals with the numerical discretization of the equations, including in \secref{sec:z_step} a discussion of the implementation of a spectral-like scheme for the advection in the direction along the guide-field:
a combination of an optimal 
third order total variation diminishing (TVD) Runge Kutta~\cite{gottlieb_total_1998} 
for the time derivative with a seventh-order upwind scheme for the 
fluxes~\cite{pirozzoli_conservative_2002}.
A series of linear and nonlinear benchmarks of the code are presented in~\secref{sec:tests}, with emphasis on Orszag-Tang-type decaying turbulence test cases.
Finally, the main points and results of this paper are summarised in ~\secref{sec:summary}. Also included for reference in Appendix A is the recently proposed modification of the KREHM model to allow for background electron temperature gradients~\cite{zocco_kinetic_2015}.
\section {Sets of Equations solved by \viriato}
\label{sec:eqs}
\viriato~solves two distinct sets of equations: (i) the Kinetic Reduced Electron
Heating Model (KREHM) equations~\cite{zocco_reduced_2011} and (ii) 
the Kinetic Reduced MHD (KRMHD) equations~\cite{schekochihin_astrophysical_2009}.
These models are briefly discussed below; we refer the interested reader to the original 
references for a detailed derivation of the equations of each model.

\subsection{The Kinetic Reduced Electron Heating Model (KREHM)}
\label{sec:krehm_eqs}
The Kinetic Reduced Electron Heating Model (KREHM) derived in Ref.~\cite{zocco_reduced_2011} is 
a rigorous asymptotic reduction of standard gyrokinetics~\cite{frieman_nonlinear_1982,howes_astrophysical_2006, garbet_gyrokinetic_2010,
krommes_gyrokinetic_2012, abel_multiscale_2012} 
applicable to plasmas that verify~\eq{KREHM_order}.
In the slab geometry that we adopt, the background magnetic field (the guide-field) is assumed to be 
straight and uniform, $\bm B_0 = B_0 \ez$.
The perturbed electron distribution function, to lowest 
order in $\sqrt{m_e/m_i}\sim\sqrt\beta_e$,
and in the gyrokinetic expansion,
is defined as 
\be
\label{delta_f_e}
\delta f_e= g_e + (\delta n_e/n_{0e}+2\vpar u_{\parallel e}/\vthe^2)F_{0e},
\ee
where $F_{0e}$ is the equilibrium Maxwellian,   
$\vthe = \sqrt{2 T_{0e}/m_e}$ is the electron thermal speed, 
$\vpar$ is the velocity coordinate parallel to the guide-field direction,
$\delta n_e$ is the electron density perturbation 
(the zeroth moment of $\delta f_e$),
and 
\be
\label{uepar}
u_{\parallel e}=-j_{\parallel}/n_{0e} e=(e/c m_e)d_e^2\nabla_\perp^2\Apar
\ee
is the parallel electron flow (the first moment of
$\delta f_e$). In this expression, $j_{\parallel}$ is the parallel current and $\Apar$ is the parallel component of the vector potential (note that, in this model, the parallel ion flow is zero to the order that is kept in the expansion); and $d_e=c/\omega_{pe}$ is the 
electron skin depth, where $\omega_{pe}=\sqrt{4\pi n_{0e} e^2/m_e}$ is the electron plasma frequency. 
All moments of $\delta f_e$ higher than $\delta n_e$ and $u_{\parallel e}$
are contained in the ``reduced'' electron distribution function 
$g_e$, e.g., parallel temperature fluctuations are given by
\be
\frac{\delta \Tpar}{T_{0e}}=\frac{1}{n_{0e}}\int d^3{\bm v}~\frac{2\vpar^2}{\vthe^2}~g_e.
\ee

For notational simplicity, let us introduce the following usual definitions:
\bea
\label{eq:ddt}
\frac{d }{d t} &=& \frac{\d}{\d t} + \frac{c}{B_0}\[\varphi, \dots\],\\
\label{eq:bdotgrad}
\hat{\bm b}\cdot\nabla &=&\frac{\d}{\d z} - \frac{1}{B_0}\[\Apar, \dots\],
\eea
where $\varphi$ is the electrostatic potential and $\[\dots,\dots\]$ denotes the Poisson bracket.
The KREHM equations are~\cite{zocco_reduced_2011}:
\begin{align}
\label{eq:mom}
\frac{1}{\noe}\frac{d \delta n_e}{d t} = -
\hat{\bm b}\cdot\nabla\frac{e}{c m_e} d_e^2\lapperp\Apar,\\
\label{eq:Ohm}
\frac{d}{dt}\(\Apar - d_e^2\lapperp\Apar\) = \eta \lapperp\Apar-
c\frac{\d\varphi}{\d z} +\frac{c T_{0e}}{e} \hat{\bm b}\cdot\nabla \(\frac{\delta n_e}{\noe} + 
\frac{\delta \Tpar}{T_{0e}}\),\\
\label{eq:ge}
\frac{d g_e}{d t} +\vpar\hat{\bm b}\cdot\nabla\(g_e-\frac{\delta \Tpar}{T_{0e}}F_{0e}\) =
C[g_e]+\(1-\frac{2\vpar^2}{\vthe^2}\)F_{0e}\hat{\bm
b}\cdot\nabla\frac{e}{cm_e}d_e^2\lapperp\Apar. 
\end{align}
Here, $\eta$ is the Ohmic diffusivity and $C[g_e]$ is the collision operator.

The perturbed electron density and the electrostatic potential are related via the 
gyrokinetic Poisson law~\cite{krommes_fundamental_2002}:
\be
\label{eq:gk_Poisson}
\frac{\delta n_e}{\noe} = \frac{Z}{\tau}\(\hat \Gamma_0 -1\)\frac{e\varphi}{T_{0e}},
\ee
where $\tau=T_{0i}/T_{0e}$ and $\hat \Gamma_0$ denotes the inverse Fourier transform of 
$\Gamma_0(\alpha)=I_0(\alpha)e^{-\alpha}$, with $I_0$ the
modified Bessel function and 
$\alpha=\kperp^2\rho_i^2/2$ ($\rhoi=v_{{\rm th} i}/\Omega_i$ is the 
ion Larmor radius, with $v_{{\rm th}i}=\sqrt{2 T_{0 i}/m_i}$ the ion thermal velocity and $\Omega_i = Ze B_0 /m_i c$ the ion gyrofrequency).

\Eq{eq:ge} is a kinetic equation for the reduced electron distribution function 
$g_e(x,y,z,\vpar,v_{\perp},t)$. An important observation is that it does not contain 
an explicit dependence on $v_{\perp}$. If such a dependence is not introduced 
by the collision operator $C[g_e]$, then $v_{\perp}$ can be integrated out, 
and the reduced electron distribution function becomes effectively 4D only, 
$g_e=g_e(x,y,z,\vpar)$.

\subsubsection{Hermite expansion}
\label{sec:KREHM_Hermite}
The use of a Hermite polynomial expansion of the distribution function is a 
well-known technique to simplify the numerical solution of kinetic equations 
such as~(\ref{eq:ge})~\cite{grant_fourier-hermite_1967, grant_transition_1967,
joyce_numerical_1971, 
knorr_plasma_1974, hammett_developments_1993, smith_dissipative_1997,sugama_collisionless_2001} --- 
see~\cite{hammett_developments_1993} in particular for an insightful discussion 
of this approach. A very convenient aspect of the Hermite formulation is 
that it enables a spectral representation of velocity space, a highly-desirable property when the available resolution is limited (as is almost invariably the case).
It is worth pointing out, furthermore, that the advantages of the Hermite representation 
transcend the numerical aspects, as it often enables one to make analytical headway 
in problems that are otherwise too complex: see, e.g., Refs.~\cite{zocco_reduced_2011, 
kanekar_fdr_2014,zocco_kinetic_2015,zocco_equivalence_2015}.  

Perhaps for both these reasons, Hermite formulations have gathered considerable interest 
recently; we refer the reader to Ref.~\cite{parker_fourier-hermite_2015} for a very comprehensive overview of recent and past work on the subject.

The Hermite expansion of $g_e$ is defined by
\be
\label{Hermite_expansion}
g_e(x,y,z,\vpar,t)=\sum_{m=2}^\infty \frac{1}{\sqrt{2^m m!}}H_m\(\frac{\vpar}{\vthe}\) 
g_m (x,y,z,t)F_{0e}(\vpar),
\ee 
where $H_m$ denotes the Hermite polynomial of order $m$ and $g_m$ is its coefficient. 
Note that $g_0=g_1=0$ because $\delta n_e$ and $u_{\parallel e}$ have been explicitly 
separated in the decomposition of $\delta f_e$ given in \eq{delta_f_e}.

Introducing this expansion into \eq{eq:ge}, and choosing a modified Lenard-Bernstein
collision operator~\cite{zocco_reduced_2011}, yields a set of coupled, 
fluid-like equations for the coefficients of the Hermite polynomials:
\begin{align}
\label{eq:gms}
\frac{d g_m}{d t} + \vthe \hat{\bm b}\cdot\nabla\(\sqrt{\frac{m+1}{2}}g_{m+1}+
\sqrt{\frac{m}{2}}g_{m-1}-\delta_{m,1}g_2\) = \nonumber\\
 -\sqrt{2}\delta_{m,2}\bhat\cdot\nabla
\frac{e}{cm_e}d_e^2\lapperp\Apar-\nu_{ei}\(m g_m-2\delta_{m,2}g_2\),
\end{align}
where $\delta_{m,2}$ is a Kronecker delta and $\nu_{ei}$ is the electron-ion 
collision frequency. In addition, this choice for $C[g_e]$ defines the resistive diffusivity:
\be
\eta\equiv\nu_{ei}d_e^2.
\ee

In the Hermite formulation, $m$ is the velocity-space equivalent of $k$ in the usual Fourier representation of position space. Thus, for example, the formation
of fine scale structures in velocity space
(as arises from phase-mixing) can be conveniently thought of as a transfer of 
energy to high $m$'s, much in the same way as the formation of fine scales 
in real space leads to energy being transferred to high wave numbers $k$ in the 
usual Fourier representation.
On the other hand, the Hermite representation introduces a closure problem, in that 
the equation for $g_m$ couples to the higher order moment $g_{m+1}$.
We shall see in \secref{sec:closure}, however, that a rigorous, nonlinear closure 
can be obtained.

\subsubsection{Reduced MHD limit}
\label{sec:RMHD_limit}
The well known reduced MHD (RMHD) equations~\cite{kadomtsev_nonlinear_1974, 
strauss_nonlinear_1976} can be obtained 
from~\eqs{eq:mom}{eq:gk_Poisson} by taking the collisional limit $\nu_{ei}\gg\omega$, 
$\kperp\ll\(\rho_{i}^{-1}, \rho_s^{-1}, d_e^{-1}\)$, where $\omega$ and $\kperp$ 
represent the typical frequencies and perpendicular wave numbers of the fluctuations, and
$\rhos = \rho_i/{\sqrt{2\tau}}$ is the ion sound Larmor radius.

In this limit, the isothermal 
approximation, $\delta \Tpar =0$, applies, and thus~\eq{eq:ge} decouples from 
\eqs{eq:mom}{eq:Ohm}. 
For $\kperp\rho_i\ll 1$, \eq{eq:gk_Poisson} becomes
\be
\label{eq:RMHD-Poisson}
\frac{\delta n_e}{n_{0e}}=\frac{1}{\Omega_i}\lapperp\Phi,
\ee
where we have defined $\Phi\equiv c \varphi / B_0$ to make contact with the 
standard terminology. Further defining $\Apar \equiv - \sqrt{4\pi n_0 m_i}\Psi$, 
we obtain:
\bea
\frac{\d}{\d t}\lapperp\Phi + \[\Phi,\lapperp\Phi\] &=&  
v_A\frac{\d}{\d z}\lapperp\Psi + \[\Psi,\lapperp\Psi\], \label{eq:vorticity}\\
\frac{\d \Psi}{\d t} + \[\Phi,\Psi\]&=& \eta \lapperp\Psi +
v_A\frac{\d \Phi}{\d z}, \label{eq:epar}
\eea
where $v_A$ is the Alfv\'en speed based on the guide-field, $
v_A=B_0/\sqrt{4\pi n_0 m_i}$.

%
\subsection{Kinetic Reduced Magnetohydrodynamics (KRMHD)}
\label{sec:slow_mode_eqs}
A different set of equations solved by \viriato~ is the 
Kinetic Reduced Magnetohydrodynamics (KRMHD) model, derived by expanding the gyrokinetic equation in terms of the small
parameter $k_\perp \rhoi$~\cite{schekochihin_astrophysical_2009}---in this sense, it is the long
wavelength limit of gyrokinetics. In this limit, the Alfv\'enic component of the turbulent
fluctuations
decouples from the compressive component. The dynamics of the system are completely
determined by the Alfv\'enic fluctuations, which are governed by the reduced MHD
\eqs{eq:vorticity}{eq:epar}. The compressive fluctuations, on the other hand, evolve
according to a kinetic equation:
\bea
\frac{d g}{dt} + \vpar \hat{\bm b}\cdot\nabla g &=& \frac{\vpar F_0}{\Lambda}\hat{\bm
b}\cdot \nabla \int d \vpar g, \label{eq:slowmodekin}
\eea
where $g$ is related to the perturbed ion distribution function [see equation (183) of
Schekochihin \etal~\cite{schekochihin_astrophysical_2009}] and
$F_0 = \exp(-\vpar^2/\vthi^2)/\sqrt{\pi \vthi}$ is a one dimensional Maxwellian. 
The parameter $\Lambda$ is a linear combination of the
physical parameters ion-to-electron temperature ratio, plasma beta, and the ion charge
[see equation (182) of Schekochihin \etal~\cite{schekochihin_astrophysical_2009}]. 

The structure of \eqref{eq:slowmodekin} is mathematically similar to that of
\eqref{eq:ge}, the main difference being that this kinetic equation is 
decoupled from the Alfv\'enic fluctuations, unlike its KREHM counterpart.

Similar to
\secref{sec:KREHM_Hermite}, one obtains the following set of equations by expanding
\eqref{eq:slowmodekin} in terms of Hermite polynomials: 
\begin{align}
    \label{eq:g0}
    &\od{g_0}{t} + \vthi \nabla_\parallel\frac{g_1}{\sqrt{2}}  = 0, \\
    \label{eq:g1}
    &\od{g_1}{t} + \vthi \nabla_\parallel\lt(g_2 + \frac{\lt(1-1/\Lambda\rt)}{\sqrt{2}}\,g_0\rt)
    = 0,\\
    \label{eq:gmeq}
    &\od{g_m}{t} + \vthi \nabla_\parallel\lt(\sqrt{\frac{m+1}{2}}\,g_{m+1} +
    \sqrt{\frac{m}{2}}\,g_{m-1}\rt) \nonumber \\
    &= C[g_m],  \quad m\ge2.
\end{align}

Notice that, unlike \eqref{eq:gms}, \eqs{eq:g0}{eq:gmeq} begin at $m=0$. Additionally,
since the term on the right hand side of \eq{eq:slowmodekin} is proportional to the first
Hermite polynomial, the
parameter $\Lambda$ makes an appearance only in the equation for $g_1$.

\section {Hermite closure}
\label{sec:closure}

The Hermite expansion transforms the original electron drift-kinetic equation, (\ref{eq:ge}), into 
an infinite, coupled set of fluid-like equations, (\ref{eq:gms}) [or, similarly for KRHMD, \eq{eq:slowmodekin} into \eqs{eq:g0}{eq:gmeq}]. 
Formally, the two representations are exactly equivalent, i.e., no information is lost by introducing the Hermite representation.
However, the numerical implementation of equations~(\ref{eq:gms}) obviously requires some form of truncation, 
i.e., given a certain number of Hermite moments, $M$, it is necessary to specify some 
prescription for $g_{M+1}$. 
In other words, as in the derivation of any fluid set of equations, the Hermite expansion introduces a closure problem.
Attempts to solve this problem have varied, from simply setting 
$g_{M+1}=0$ (e.g., \cite{grant_fourier-hermite_1967,hammett_developments_1993,
gibelli_spectral_2006,parker_fourier-hermite_2015}), to polynomial closures in which $g_{M+1}$ is extrapolated 
from a number of previous moments~\cite{knorr_plasma_1974,gibelli_spectral_2006}. 
Particularly noteworthy is 
the approach followed by Hammett and 
co-workers~\cite{hammett_fluid_1990, hammett_developments_1993, dorland_gyrofluid_1993,
smith_dissipative_1997,snyder_gyrofluid_2001} where closures have been carefully designed to rigorously 
capture the linear Landau damping rates (as well as gyro-radius effects and 
dominant nonlinearities).

In the system of equations under consideration here, it turns out that an 
asymptotically exact closure can be obtained in the large $M$ limit. 
Let us consider that the collision frequency is small but finite. 
Then, there will be a range of $m$'s for which the collisional term is negligible ---
one may think of this as the $m$ inertial range: energy is injected into low $m$'s via the coupling with Ohm's law, and cascades (phase-mixes) to higher $m$'s. However, as $m$ increases, a dissipation range is encountered, when the collisional term in \eq{eq:gms} [or in \eq{eq:gmeq}] is no longer subdominant with respect to the other terms. 
Roughly speaking, in the dissipation range, energy arrives at $g_m$ from $g_{m-1}$ and is mostly dissipated there; only an exponentially smaller fraction is passed on to $g_{m+1}$.
One thus expects that $g_{m+1}/g_{m}\ll 1$ in the dissipation range, by definition. 
The implication of this is that, for $m=M$ in the dissipation range, the dominant balance in the equation for $g_{M+1}$ must be
\be
\label{eq:closure}
\vthe \hat{\bm b}\cdot\nabla \sqrt{\frac{M+1}{2}}g_{M}\approx
-\nu_{ei}(M+1) g_{M+1}.
\ee
Solving this equation for $g_{M+1}$ yields the sought closure~\cite{loureiro_fast_2013,zocco_kinetic_2015}.
The equation for $g_M$ therefore becomes:
\be
\label{eq:last_g}
\frac{d g_M}{d t} 
-\kappa_{\parallel e} \hat{\bm b}\cdot\nabla\(\hat{\bm b}\cdot\nabla g_M\)
+\vthe \hat{\bm b}\cdot\nabla\sqrt{\frac{M}{2}}g_{M-1} =
 -\nu_{ei}M g_M,
\ee
where $\kappa_{\parallel e}\equiv \vthe^2/2\nu_{ei}$ is the parallel (Spitzer) thermal diffusivity\footnote{Note that if one wishes to close the system at $M=2$ (i.e., the semi-collisional limit), then this 
equation needs to include the term proportional to the electron current [the first term on the RHS of \eq{eq:gms}], becoming equation (99) of Ref.~\cite{zocco_reduced_2011}.}. 
It is easy to see how the exact same reasoning leads to the equivalent closure for \eq{eq:gmeq}.

It can be useful to have an {\it a priori} estimate of the value of $M$ required to formally justify the asymptotic closure, for a given collision frequency. One such {\it linear} estimate is provided in Ref.~\cite{zocco_reduced_2011}: if the Hermite spectrum is in steady-state, then the collisional cutoff, $m=m_c$, can be shown to occur at\footnote{This discussion implicitly assumes that one is dealing with a turbulent situation in statistical steady state. Alternatively, one may wish to analyse a linear instability; in that case, another cutoff appears, $m_\gamma=(|k_\parallel |\vthe/(2\sqrt{2}\gamma)^2$~\cite{zocco_reduced_2011}. If $m_\gamma<m_c$ then the collisional cutoff is superseded. This does not affect any of the considerations drawn here.}:
\be
\label{m_cutoff}
m_c=\(\frac{3}{2\sqrt{2}}\frac{|k_{\parallel}|\vthe}{\nu_{ei}}\)^{2/3}.
\ee
Thus, we expect the Hermite closure, \eq{eq:closure}, to be valid if $M\gg m_c$.

The numerical implementation of~\eq{eq:last_g} introduces some difficulties and will be discussed in \secref{sec:closure_implementation}.
\subsection{Hypercollisions}
Since our primary interest lies in weakly collisional plasmas, one finds that $m_c\gg1$.
For example, a simple estimate using standard parameters for the solar corona suggests $m_c\approx 10^4$; certain experiments on JET~\cite{saibene_h-mode_2007} suggest $m_c\approx 180$ in the edge region, considerably smaller than for the solar corona, but still quite large. Further noticing that such cases are invariably tied to a broad range of spatial scales, thereby also requiring high spatial resolutions, renders obvious the impracticability of such computations: not only 
must one solve a very large set of nonlinear, coupled PDE's, as 
also the stiffness increases, due to the coefficients proportional 
to $\sqrt{m}$.
One possibility of avoiding this problem is to artificially enhance the value of the collision frequency. 
Note however that $m_c\sim \nu_{ei}^{-2/3}$, i.e., a relatively weak scaling, implying that cutting the number of necessary $m$'s down to computationally manageable sizes would require drastic increases in the collision frequency. To make matters worse, the
collision operator scales only linearly with $m$, implying that in fact one needs to retain $m\gg m_c$ to adequately capture the dissipation range and validate the closure.

One way to circumvent these difficulties is to make use of a `hyper-collision' operator, i.e., add 
a term of the form $-m^{h}\nu_H g_m$ to the RHS of \eq{eq:gms}. Here, $h$ 
is the order of the hyper-diffusion operator (a typical value would be $h=6$)
and $\nu_H$ is a numerically-based coefficient defined such that energy arriving at $m=M$ can be dissipated in one timestep:
\be
\label{hyper-coeff}
\frac{\d g_M}{\d t}\sim M^{h}\nu_H g_M.
\ee
Thus, in practice, one may simply set~\cite{borue_spectra_1997,loureiro_fast_2013}:
\be
\nu_H=1/(\dt M^{h}).
\ee

It is worth remarking that if it is possible to choose a value of $M$ that is very deep into the dissipation range, then presumably the issue of which closure to implement becomes less sensitive, and it may be justified to simply set $g_{M+1}=0$.
Indeed, we have performed simulations with both closures and observed no differences (not reported in this paper).

Finally, we point out that in alternative to a hyper-collision operator one may use a spectral filter (in $m$-space), such as the one of Hou and Li~\cite{hou_computing_2007}, as proposed by Parker and Dellar~\cite{parker_fourier-hermite_2015} (see also~\secref{sec:Hou_Li} for a discussion of this filter in Fourier space).

\section{Energy}
\label{sec:energy}
In the absence of collisions, \eqs{eq:mom}{eq:ge} conserve a quadratic invariant usually referred to as free energy~\cite{schekochihin_astrophysical_2009}. 
This quantity can be defined as $W=W_{\rm fluid}+H_e$, where~\cite{zocco_reduced_2011}
\be
W_{\rm fluid}=\sum_{\bm k} \[1+\frac{1}{\tau}(1-\Gamma_0)\]\frac{1}{\tau}(1-\Gamma_0)
\frac{e^2n_{0e}|\varphi_{\bm k}|^2}{2T_{0e}}
+\int\frac{d^3{\bm r}}{V}\frac{|\nabla_\perp \Apar|^2 + d_e^2|\nabla^2_\perp\Apar|^2}{8\pi}
\ee
is the ``fluid'' (electromagnetic) part of the free energy, and 
\be
H_e = \int \frac{d^3{\bm r}}{V}\int d^3{\bm v}\frac{T_{0e}g_e^2}{2F_{0e}}
\ee
is the electron free energy (i.e., the free energy associated with the reduced electron distribution function $g_e$).

Upon introducing the Hermite expansion of $g_e$, \eq{Hermite_expansion}, and allowing for finite collisions (modelled by the Lenard-Bernstein collision operator), one finds that $W$ evolves according to the following equation~\cite{zocco_reduced_2011}:
\be
\label{eq:energy}
\frac{d}{dt}W_{\rm fluid} + \frac{d}{dt}\int \frac{d^3{\bm r}}{V}\frac{n_{0e}T_{0e}}{2}
\sum_{m=2}^{\infty}g_m^2 = 
-n_{0e}T_{0e}\nu_{ei}\int\frac{d^3{\bm r}}{V}\sum_{m=3}^\infty m g_m^2
-\frac{4\pi}{c^2}\eta\int\frac{d^3{\bm r}}{V}j_\parallel^2.
\ee

The above equation is exact. However, as discussed in \secref{sec:closure}, the numerical implementation of the Hermite expansion requires that only a finite number of Hermite polynomials are kept, and some form of closure to the expansion is required. If we adopt the closure described by \eq{eq:closure}, and truncate the expansion at $m=M$, \eq{eq:energy} adopts the truncated form:
\bea
\label{eq:trunc_energy}
&&\frac{d}{dt}W_{\rm fluid} + \frac{d}{dt}\int \frac{d^3{\bm r}}{V}\frac{n_{0e}T_{0e}}{2}
\sum_{m=2}^{M}g_m^2 = \nonumber\\
&&-n_{0e}T_{0e}\nu_{ei}\int\frac{d^3{\bm r}}{V}\sum_{m=3}^M m g_m^2
-n_{0e}T_{0e}\kappa_e \int\frac{d^3{\bm r}}{V}\(\hat{\bm b}\cdot\nabla g_M\)^2
-\frac{4\pi}{c^2}\eta\int\frac{d^3{\bm r}}{V}j_\parallel^2,\nonumber\\
\eea
where the second term on the RHS is due to the specific closure that we have used (and would vanish if, for example, we instead use the simpler closure $g_{M+1}=0$.)

The same arguments that were invoked to motivate the Hermite closure in \secref{sec:closure} apply here to justify the asymptotic equivalence of the full form of the energy balance, \eq{eq:energy} and its truncated version, \eq{eq:trunc_energy} --- that is, as long as $M$ is as large as required for $g_M$ to lie in the collisional (i.e., $m$-) dissipation range, one expects the terms neglected in going from \eq{eq:energy} to \eq{eq:trunc_energy} to be exponentially small. 

The corresponding equation for KRMHD is equation (4.7) of Ref.~\cite{kanekar_fdr_2014}. The closure that we propose in ~\secref{sec:closure} can be implemented in this set of equations in a similar way, and it is straightforward to obtain the KRMHD counterpart of \eq{eq:trunc_energy}. 
\section{Normalizations}
\label{sec:KREHM_norms}
The normalizations that we adopt for the KREHM set of equations (\ref{eq:mom},\ref{eq:Ohm},\ref{eq:gk_Poisson},\ref{eq:gms}) are:
\begin{itemize}
\item{\it Length scales:}
  \be
  (\hat x,\hat y)= (x,y)/\Lperp;\quad \hat z = z/\Lpar,
  \ee
  where $\Lperp,~\Lpar$ are, respectively, the perpendicular and parallel (to the guide-field) 
  reference length-scales.
\item{\it Times:}
  \be
  \hat t = t/\tA,
  \ee
  where $\tA=\Lpar/v_A$ is the parallel Alfv\'en time.
\item{\it Fields:}
  \bea
  (\hat n_e, \hat g_m) &=& \tA \Omega_i\(\frac{\delta n_e}{\noe},g_m\),\\
  \hat \varphi &=& \frac{c}{B_0}\frac{\tA}{\Lperp^2}\varphi,\\
  \hat \Apar &=& \frac{\Lpar}{\Lperp}\frac{\Apar}{\Lperp B_0}.
  \eea
\end{itemize}
Under these normalizations, equations~(\ref{eq:mom}),~(\ref{eq:Ohm}) and (\ref{eq:gms}) become:
\bea
\label{momentum}
&&\frac{d n_e}{d t} =  \[\Apar, \lapperp\Apar\] -\frac{\d}{\d z}\lapperp\Apar,\\
\label{Ohm}
&&\frac{d}{d t}\(\Apar - d_e^2\lapperp\Apar\)  
= \eta\lapperp\Apar + 
\rho_s^2\[n_e+\sqrt{2}g_2,\Apar\] \nonumber\\
&&\qquad\qquad -\frac{\d \varphi}{\d z} + \rhos^2\frac{\d}{\d z}\(n_e + \sqrt{2} g_2\),\\
\label{g2_eq}
&&\frac{d g_2}{d t} =
\sqrt{3}\frac{\rho_s}{d_e}\left\{ \[\Apar,g_3\]-\frac{\d g_3}{\d z}\right\} 
+\sqrt{2} \left\{\[\Apar,\lapperp\Apar\]-\frac{\d}{\d z}\lapperp \Apar\right\}, \\
\label{gm_eq}
&&\frac{d g_m}{d t} = 
\sqrt{m+1}\frac{\rho_s}{d_e} \left\{\[\Apar,g_{m+1}\] -\frac{\d g_{m+1}}{\d z}\right\}
+\sqrt{m}\frac{\rho_s}{d_e} \left\{\[\Apar,g_{m-1}\] -\frac{\d g_{m-1}}{\d z}\right\}\nonumber\\
&&\qquad\qquad- m\nu_{ei}g_m,  \quad m>2.
\eea
where now
\be
\frac{d}{dt} = \frac{\d }{\d t} + \[\varphi,\dots\].
\ee
The normalized form of the quasi-neutrality \eq{eq:gk_Poisson} is
\be
\label{Poisson}
n_e=\frac{2}{\rho_i^2}\[\hat\Gamma_0(\alpha)-1\]\varphi.
\ee
It can immediately be seen that neglecting the $g's$ reduces the above set of 
equations to the simpler two-field gyrofluid model treated 
in~\cite{loureiro_iterative_2008}.

For the KRMHD set of \eqs{eq:g0}{eq:gmeq} the normalisation of space and time are as above, upon which the only modification is the conversion of the prefactor $\vthi$ into $\sqrt{\beta_i}$, where $\beta_i=8\pi n_{0i} T_{0i}/B_0^2$ is the ion plasma beta. The normalisation of the Hermite moments $g_m$ is arbitrary since those equations are linear in $g_m$.

\section{Numerical discretization}
\label{sec:numerics}
The RHS of \eqs{momentum}{gm_eq} is conveniently separated into operators acting either in the direction perpendicular ($x,~y$) or parallel ($z$) to the guide field. 
This suggests that an efficient way of integrating those equations  is to use operator splitting techniques such as to individually handle each class (perpendicular or parallel) of operators.
\viriato~allows for both Godunov~\cite{godunov_1959} or Strang splitting~\cite{strang_1968}. Although Godunov splitting is formally only 1st-order accurate, direct comparisons of both splitting schemes performed by us (not reported here) yield undistinguishable results. Thus, by default, \viriato~employs Godunov splitting (as it is computationally cheaper); all results reported in \secref{sec:tests} are obtained with this option.

We now detail the algorithms employed for the perpendicular and parallel steps.

\subsection{Perpendicular direction}
The numerical discretisation of \eqs{momentum}{gm_eq} is the 
straightforward generalisation of that derived in~\cite{loureiro_iterative_2008}\footnote{With the exception that here we do not include the semi-implicit operator that was the main subject of Ref.~\cite{loureiro_iterative_2008}. Although the semi-implicit operator derived there can easily be extended to the KREHM equations --- by using the full kinetic Alfv\'en wave dispersion relation, \eq{eq:LKAW} --- this is not the focus of this paper and we prefer to leave it out of the discussion.}. 
For presentational simplicity, let us denote the nonlinear terms (i.e., 
the Poisson brackets) in~\eqs{momentum}{gm_eq} by generalised operators, 
such that we have\footnote{We include here also, in Ohm's law, an external electric field $E_{\rm ext}=-\eta \lapperp A_{\parallel,eq}$ which is used in tearing mode simulations to prevent the resistive diffusion of the background (reconnecting) magnetic field.}:
\bea
\frac{\d n_e}{\d t} &=& \NN(n_e,\Apar),\\
(1+\kperp^2d_e^2)\frac{\d \Apar}{\d t} &=& \mathcal A(n_e,\Apar, g_2) - 
\eta\kperp^2(\Apar-A_{\parallel,eq}),\\
\frac{\d g_2}{\d t}&=&\GG_2(n_e,\Apar,g_2,g_3),\\
\frac{\d g_m}{\d t}&=&\GG_m(n_e,\Apar,g_{m-1},g_m,g_{m+1}) -m\nu_{ei}g_m.
\eea
Then, the integration scheme is as follows. First we take a predictor step:
\bea
n_e^{n+1,*} &=& n_e^n + \dt \NN(n_e^n,\Apar^n),\\
\Apar^{n+1,*} &=& e^{-D_\eta\dt}\Apar^n + \(1-e^{-D_\eta\dt}\)A_{\parallel,eq} + \nonumber\\
&&\frac{\dt}{2}\frac{1+e^{-D_{\eta}\dt}}
{1+\kperp^2d_e^2}\AA(n_e^n,\Apar^n,g_2^n),\\
g_2^{n+1,*} &=& g_2^n + \dt \GG_2(n_e^n,\Apar^n,g_2^n,g_3^n),\\
g_m^{n+1,*} &=& e^{-m\nu_{ei}\dt}g_m^n + \nonumber\\ 
&&\frac{\dt}{2}\(1+e^{-m\nu_{ei}\dt}\)\GG_m(n_e^n,\Apar^n,g_{m-1}^n,g_m^n,
g_{m+1}^n),
\eea
where $\deta=\kperp^2\eta/(1+\kperp^2d_e^2)$. This is followed by the 
corrector step, which can be iterated $p$ times until the desired level of convergence is achieved:
\bea
\Apar^{n+1,p+1} &=& e^{-D_\eta\dt}\Apar^n + 
\(1-e^{-D_\eta\dt}\)A_{\parallel,eq} +
\frac{\dt}{2}\frac{e^{-D_{\eta}\dt}}
{1+\kperp^2d_e^2}\AA(n_e^n,\Apar^n,g_2^n) +\nonumber\\
&&\frac{\dt}{2}\frac{1}{1+\kperp^2d_e^2}
\AA(n_e^{n+1,p},\Apar^{n+1,p},g_2^{n+1,p}),\\ 
%
n_e^{n+1,p+1} &=& n_e^n + \frac{\dt}{2} \NN\(n_e^n,\Apar^n\) + 
\frac{\dt}{2}\NN\(n_e^{n+1,p},\Apar^{n+1,p+1}\),\\
g_2^{n+1,p+1} &=& g_2^n + 
\frac{\dt}{2}\GG\(n_e^n,\Apar^n,g_2^n,g_3^n\) + \nonumber\\
&&\frac{\dt}{2}\GG_2(n_e^{n+1,p+1},\Apar^{n+1,p+1},g_2^{n+1,p},g_3^{n+1,p}),\\
g_m^{n+1,p+1} &=& e^{-m\nu_{ei}\dt}g_m^n + 
\frac{\dt}{2}e^{-m\nu_{ei}\dt}\GG_m\(n_e^n,\Apar^n,g_{m-1}^n,g_m^n,
g_{m+1}^n\) + \nonumber\\
&&\frac{\dt}{2}\GG_m\(n_e^{n+1,p+1},\Apar^{n+1,p+1},
g_{m-1}^{n+1,p+1},g_m^{n+1,p},g_{m+1}^{n+1,p}\).
\eea

For presentational simplicity, we have not included here the hyper-diffusion and hyper-collisions operators, but it is trivial to do so: they are handled in the same way as the resistivity or the collisions are in the above equations.
\subsubsection{Dealiasing {\it vs.} Fourier smoothing}
\label{sec:Hou_Li}
To deal with the possibility of aliasing instability~\cite{boyd_chebyshev_2001}, 
\viriato~offers two options. One is the standard $2/3$'s rule~\cite{orszag_elimination_1971}, 
where the Fourier transformed fields are multiplied by a step function $\rho(k/\kmax)$ defined by:
\be
\label{two_thirds_rule}
\rho(k/\kmax)=\begin{cases}
1& \text {if $|k|/\kmax\le 2/3$}, \\
0& \text{if $|k|/\kmax> 2/3$},
\end{cases}
\ee
where $\kmax=N/2$ for a grid with $N$ points.
The second option is the high-order Fourier filter proposed by 
Hou \& Li~\cite{hou_computing_2007}:
\be
\label{hou_li_filter}
\rho(k/\kmax)=\exp\[-36\(|k|/\kmax\)^{36}\].
\ee
Compared to~\eq{two_thirds_rule}, the Hou-Li filter retains 12-15\% more active Fourier modes 
in each direction. For other advantages of this filter, and justification of
its specific functional form, the reader is referred to Ref.~\cite{hou_computing_2007}.
Tests reported in Refs.~\cite{hou_computing_2007, grafke_numerical_2008, hou_blow-up_2009, 
dellar_lattice_2013} unanimously 
confirm the numerical superiority of the Hou-Li filter over the $2/3$'s rule dealiasing, 
as will our results presented in \secref{sec:OT}.

\subsection{Parallel direction}
\label{sec:z_step}
\viriato~has inbuilt two distinct methods for the integration of the equations in the direction along the guide-field, $z$: 
a MacCormack scheme~\cite{MacCormack}, 
and a combination of a third-order total variation diminishing (TVD) Runge Kutta method for 
the time derivative~\cite{gottlieb_total_1998} with a 
seventh-order upwind discretization for the fluxes~\cite{pirozzoli_conservative_2002} 
(TVDRK3UW7 for short).
The MacCormack scheme is fairly standard (see, e.g., ~\cite{jardin_computational_2010,
durran_numerical_2010} for 
textbook presentations) 
and there is no need 
to detail it here. The TVDRK3UW7 is not as conventional and is described below.

\subsubsection{Characteristics}  
\label{sec:charact}
The $z$-advection step consists in solving the following set of equations:
\be     
\label{z_system}
\frac{d \bf u}{d t} = A \frac{d \bf u}{d z}, 
\ee
where 
\be
{\bf u}=(n_e, \Apar, g_2, ..., g_M)^T 
\ee
is the solution vector 
and $A$ is tridiagonal matrix of size $(M+1)\times (M+1)$ whose 
only non-zero entries are the coefficients of the $z$-derivatives, as follows:
\bea
&&A_{k,k+1} = \left\{\kperp^2, \sqrt{2}\rhos^2, -\sqrt{3}\frac{\rho_s}{d_e},\cdots, 
-\sqrt{m+1}\frac{\rho_s}{d_e},\cdots\right\}, \nonumber\\
&&\quad k=1,\cdots, M,\\
&&A_{k,k-1} = \left\{\frac{1}{1+\kperp^2d_e^2}\(\rho_s^2-\frac{\rho_i^2}{2(\Gamma_0-1)}\), 
\sqrt{2}\kperp^2, -\sqrt{3}\frac{\rho_s}{d_e},\cdots, -\sqrt{m}\frac{\rho_s}{d_e},\cdots\right\}, \nonumber\\
&&\quad k=2,\cdots, M.
\eea

To be able to use upwind schemes, 
we need to write \eq{z_system} in characteristics form, i.e., 
we need to diagonalize $A$. 
To do so, we introduce the matrix $P$ such that \eq{z_system} becomes
\be    
P^{-1} \frac{d\bf u}{dt} = P^{-1} A P P^{-1} \frac{d\bf u}{dz}.
\ee
We define ${\bf w} \equiv P^{-1}{\bf u}$ and solve for $P$ requiring that 
\be    
P^{-1}AP = D, 
\ee    
where $D$ is a diagonal matrix.      
The equation for $\bf w$ is now in characteristics form:
\be    
\label{z_charact}
\frac {d\bf w}{d t}  = D \frac{d\bf w}{dz}, 
\ee
namely, if $D(j)>0$, $w_j$ is a right propagating wave field, and vice-versa.
Finally, since the entries of $A$ are independent of $z$, so are the entries of $D$.
\Eq{z_charact} can thus be written in flux-conservative form:
\be
\label{z_fluxform}
\frac {d\bf w}{d t}  = \frac{d\bf F}{dz}, 
\ee
where ${\bf F}\equiv D{\bf w}$.
    
As is well known from standard linear algebra, the diagonal entries of the matrix $D$ 
are the eigenvalues of $A$, whereas $P$ is the matrix whose column vectors
are the eigenvectors of $A$. 
In \viriato, both eigenvalues and eigenvectors of $A$ are easily obtained with 
the linear algebra package LAPACK~\cite{lapack_1999}.

As an example, let us consider the simplest possible case: the reduced-MHD limit. 
Matrix $A$ becomes:
\be
A=\begin{bmatrix}
0 & \kperp^2\\
\frac{1}{\kperp^2} & 0
\end{bmatrix}.
\ee
It is a trivial exercise to obtain the matrices $P$, $P^{-1}$ and $D$. They are:
\be
P=\begin{bmatrix}
-\kperp^2 & \kperp^2\\
1 & 1
\end{bmatrix}, \quad
P^{-1}=\begin{bmatrix}
-\frac{1}{2\kperp^2} & \frac{1}{2}\\
\frac{1}{2\kperp^2} & \frac{1}{2}
\end{bmatrix},\quad
D=\begin{bmatrix}
-1 & 0\\
0 & 1
\end{bmatrix}.
\ee

In this case, the characteristic fields are
\be 
{\bf w} = P^{-1}{\bf u}=\frac{1}{2}(\Apar-\frac{n_e}{\kperp^2}, \Apar+\frac{n_e}{\kperp^2})^T.
\ee
To relate this to a more familiar case, note that, using \eq{Poisson} in the $\kperp\rhoi\ll 1$ limit to express the electron density perturbation in terms of the electrostatic potential, $n_e = -\kperp^2\varphi$, 
we immediately recognize the commonly used Elsasser potentials:
\be
w^{\pm} = \frac{1}{2}(\Apar\pm\varphi).
\ee

Note that the entries of $A$ are constants, independent of either time or space. Thus, 
the matrices $P$, $P^{-1}$ and $D$ need only to be calculated once per run, with 
negligible impact on the overall code performance.

\subsubsection{Fluxes}
\label{sec:fluxes}
The derivative of the flux $\bf F$ is computed using a seventh-order upwind 
scheme~\cite{pirozzoli_conservative_2002}:
\be
\(\frac{d{\bf F}}{dz}\)_i=\frac{{\bf F}_{i+1/2}-{\bf F}_{i-1/2}}{\Delta z},
\ee
where, for the $j$th component of $\bf F$, we have
\bea
\label{flux1}
F_{i+1/2}^j=-\frac{1}{140}F_{i+4}^j+\frac{5}{84}F_{i+3}^j-\frac{101}{420}F_{i+2}^j
+\frac{319}{420}F_{i+1}^j\nonumber\\
+\frac{107}{210}F_{i}^j-\frac{19}{210}F_{i-1}^j+\frac{1}{105}F_{i-2}^j, \quad D_j>0,
\eea
\bea
\label{flux2}
F_{i+1/2}^j=-\frac{1}{140}F_{i-3}^j+\frac{5}{84}F_{i-2}^j-\frac{101}{420}F_{i-1}^j
+\frac{319}{420}F_i^j\nonumber\\
+\frac{107}{210}F_{i+1}^j-\frac{19}{210}F_{i+2}^j+\frac{1}{105}F_{i+3}^j, \quad D_j<0.
\eea

\subsubsection{Time derivative}
\label{sec:time_der}
For the time integration of \eq{z_fluxform} we follow~\cite{samtaney_numerical_2012}. 
The time derivative is discretized using an optimal third-order total variation diminishing (TVD) 
Runge Kutta method~\cite{gottlieb_total_1998}:
\bea
\label{TVDRK}
&&{\bf w}^{(1)} = {\bf w}^{(n)} + \dt \frac{d{\bf F}^{(n)}}{dz},\nonumber\\
&&{\bf w}^{(2)} = \frac{3}{4}{\bf w}^{(n)} + \frac{1}{4}{\bf w}^{(1)} +
\frac{1}{4}\dt \frac{d{\bf F}^{(1)}}{dz},\\
&&{\bf w}^{(n+1)} = \frac{1}{3}{\bf w}^{(n)} + \frac{2}{3}{\bf w}^{(2)} +
\frac{2}{3}\dt \frac{d{\bf F}^{(2)}}{dz}.\nonumber
\eea

The final step is to compute ${\bf u}^{(n+1)}=P{\bf w}^{(n+1)}$.

Compared to the MacCormack method, the TVDRK3UW7 scheme just described has the disadvantage 
of being somewhat slower, as it 
requires three evaluations of the right hand side (as opposed to only two for MacCormack) 
and there are more communications involved between different processors to compute the fluxes, \eqs{flux1}{flux2}.
This is partially offset by the fact that the TVDRK3UW7 scheme  requires much fewer grid points per 
wavelength than the MacCormack method for an adequate resolution, as will be exemplified in \secref{sec:maccormack_vs_tvdrk3uw7}.

\subsection{Numerical implementation of the Hermite closure}
\label{sec:closure_implementation}
Expanding the $\hat{\bm b}\cdot\nabla$ operator in the closure term in \eq{eq:last_g}, we find that it becomes:
\bea
\label{eq:gM}
&&\frac{d g_M}{d t}= 
-\vthe \hat{\bm b}\cdot\nabla\sqrt{\frac{M}{2}}g_{M-1}\nonumber\\
&&+\kappa_{\parallel e} \left\{\frac{\d^2 g_M}{\d z^2}-\frac{1}{B_0}\frac{\d}{\d z}\[\Apar,g_M\]-\frac{1}{B_0}\[\Apar,\frac{\d g_M}{\d z}\] + \frac{1}{B_0^2}\[\Apar,\[\Apar,g_M\]\]\right\}\nonumber\\
&& -\nu_{ei}M g_M.
\eea

As we have discussed in previous sections, the numerical algorithm employed in \viriato~uses operator splitting methods to deal separately with the $z$-derivatives and with the Poisson brackets  (i.e., it splits the dynamics parallel and perpendicular to the magnetic guide-field). This raises a difficulty when discretising the equation above, which contains mixed terms (the second and third terms inside the curly brackets) introduced by the closure, \eq{eq:closure}; this is an especially subtle issue when the $z$-step scheme advects the equations in characteristics form, as is the case of the TVDRK3UW7 that we employ (and would equally be the case for any other upwind scheme).

Simple solutions to this problem require abandoning the operator splitting scheme and forsaking the use of the characteristics form for the $z$-derivative terms of the equations, both of which are not only highly convenient from the point of view of numerical accuracy and stability, but also physically motivated. One possibility would be to treat this equation differently from all other equations solved by the code. Although this is certainly possible, at this 
stage we have chosen not to introduce this additional complexity. As such, the actual form of \eq{eq:gM} implemented in \viriato~is
\bea
\label{eq:gM_closed}
\frac{d g_M}{d t}&= &
-\vthe \hat{\bm b}\cdot\nabla\sqrt{\frac{M}{2}}g_{M-1}
+\kappa_{\parallel e} \left\{\frac{\d^2 g_M}{\d z^2} + \frac{1}{B_0^2}\[\Apar,\[\Apar,g_M\]\]\right\}\nonumber\\
&& -\nu_{ei}M g_M.
\eea

We emphasize that the dropping of the mixed terms is purely for algorithmic reasons. From the physical point of view those terms are, {\it a priori}, as important as the closure terms which are kept; their implementation is thus left to future work. A serious drawback of this approach, for example, is that the semi-collisional limit of the KREHM equations (which results from setting $M=2$, see Section V.C of Ref.~\cite{zocco_reduced_2011}) is, therefore, not correctly captured.

On the other hand, note that:
(i) for 2D problems, our implementation of the closure is exact;
(ii) for simple linear 3D problems [where the background magnetic field is simply
given by that guide-field (which is the setup used to investigate Alfv\'en wave propagation in \secref{sec:KAW}), the numerical implementation of the closure is also exact;
(iii) in weakly collisional plasmas (which are our main focus), provided that $M$ is sufficiently large to lie in the collisional dissipation range, one expects $g_{M+1}\ll g_M$ and thus the actual functional form of the closure may not be very important;
(iv) if we first apply the operator splitting scheme (i.e., the separation of the perpendicular and parallel operators) and then impose our closure scheme on the parallel and perpendicular equations separately, we would obtain \eq{eq:gM_closed} instead of \eq{eq:gM}.

Finally, we remark that adopting \eq{eq:gM_closed} as the evolution equation for $g_M$ changes the second term on the RHS of the energy balance equation, (\ref{eq:trunc_energy}), in the obvious way.

\section{Numerical tests}
\label{sec:tests}
In this section, we report an extensive suite of linear and nonlinear benchmarks 
of \viriato.

\subsection{Comparison of the MacCormack and the TVDRK3UW7 methods}
\label{sec:maccormack_vs_tvdrk3uw7}
To illustrate the relative merits of the two numerical schemes for the $z$-advection available 
in \viriato, we carry out a simple test in the limit of isothermal electrons and cold ions.
\Eqs{momentum}{gm_eq} and \eq{Poisson} reduce to
\bea
\label{iso_cold_ne}
\frac{\d n_e}{\d t}&=&\kperp^2\frac{\d \Apar}{\d z},\\
\label{iso_cold_Apar}
\frac{\d \Apar}{\d t} &=& \frac{1}{\kperp^2}\frac{1+\kperp^2\rhos^2}{1+\kperp^2d_e^2}
\frac{\d n_e}{\d z}.
\eea
The initial condition we adopt is:
\be
\Apar(z,t=0) = \frac{\tanh\[k(z+0.25)\]+\tanh\[k(z-0.25)\]}{2}.
\ee
\Eqs{iso_cold_ne}{iso_cold_Apar} are solved on a periodic box $-L\le z \le L$, with $L=\pi$. 
The grid step size is 
$\Delta z = 2L/64$. The time step is set by the CFL condition $\dt=0.25 \Delta z/v$, where 
$v=\sqrt{(1+\kperp^2\rhos^2)/(1+\kperp^2d_e^2)}$. We chose $\kperp=1$, $\rhos=1$ and $d_e=0.01$.
There is no explicit dissipation in this test.

A measure of how well resolved the wave front is is given by the parameter 
$\hat k = k\Delta z = 2\pi/n_p$, where $n_p$ is the number of grid points per wavelength.
We test the behaviour of the MacCormack and TVDRK3UW7 schemes for three representative values
of $\hat k = 0.3,~1,~3$ (note that the highest resolvable wave number corresponds to 
$n_p=2$, i.e., $\hat k = \pi$). For each of these cases, the equations are integrated for 
10 transit times across the box, $t_{\rm transit}=2L/v$. 

\begin{figure}
  \center
  \includegraphics[width=9cm]{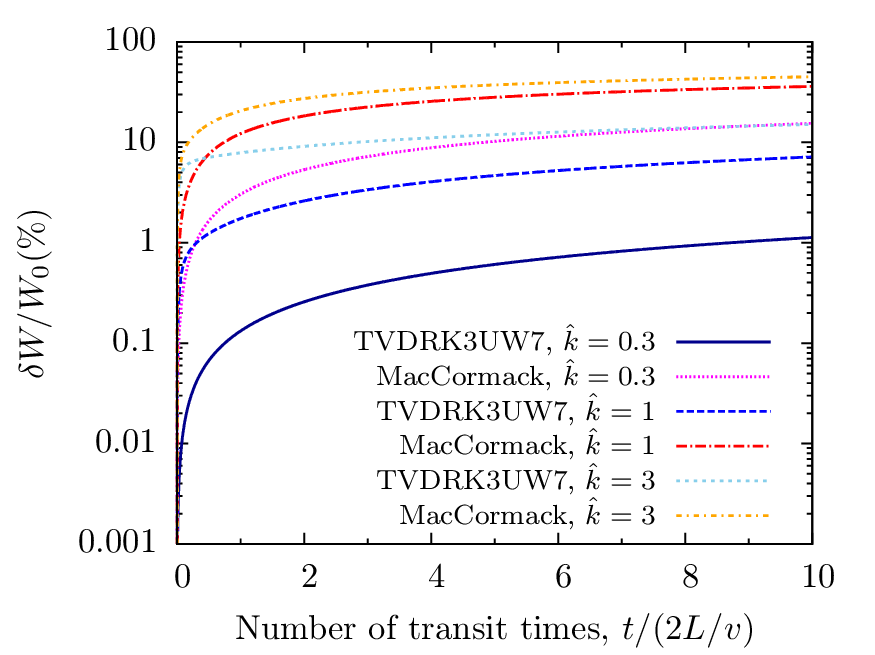}
  \caption{Energy conservation for the MacCormack and the TVDRK3UW7 schemes for the linear 
advection test problem defined in \Eqs{iso_cold_ne}{iso_cold_Apar}. The $x$-axis 
is the time normalized by the transit time across the simulation box, $2L/v$. 
The $y$-axis is the variation in energy ($\delta W$) normalized by the initial energy, $W_0$.
The parameter $\hat k = k\Delta z = 2\pi/n_p$, where $n_p$ is the number of grid points per 
wavelength.
}
  \label{fig:energy_cons}
\end{figure}

Time traces of the energy conservation for both schemes are plotted in \fig{fig:energy_cons}.
As expected, the TVDRK3UW7 scheme behaves remarkably better than MacCormack.
Notice, for example, that for the extreme case of $\hat k=3$, TVDRK3UW7 yields an amount of energy
loss after 10 crossing times of $\sim 15\%$, very similar to what is obtained with the 
MacCormack scheme for the ten times better resolved case of $\hat k=0.3$.

\begin{figure}
  \center
  \includegraphics[width=12cm]{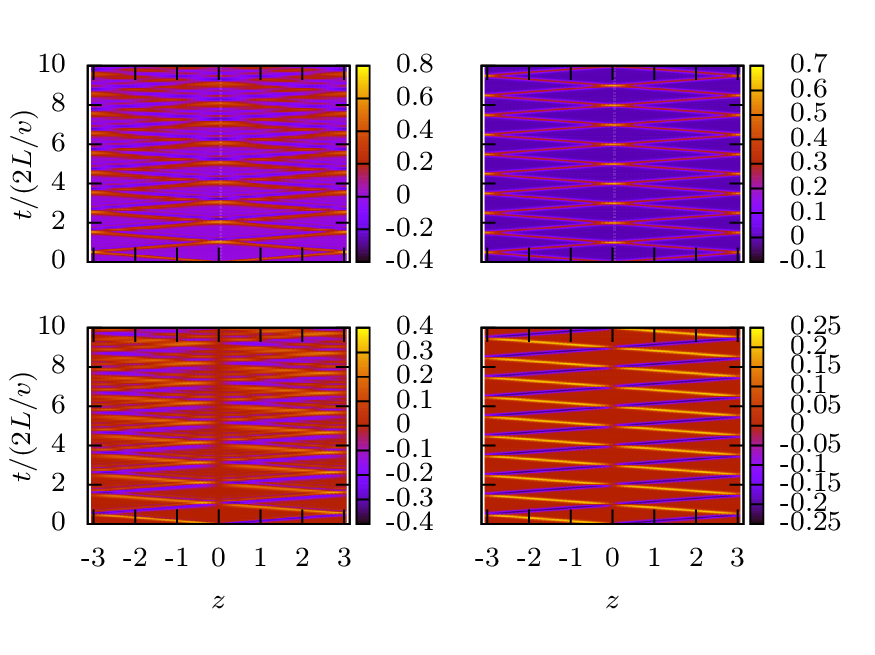}
  \caption{Results from the linear advection test problem of \secref{sec:maccormack_vs_tvdrk3uw7}. 
Contour plots of the time evolution of $\Apar$ (top) and $n_e$ (bottom) using the 
MacCormack scheme (left panels) and the TVDRK3UW7 scheme (right panels), for the case $\hat k=1$.
The MacCormack scheme is seen to introduce strong Gibbs oscillations, which are remarkably
minimized by the TVDRK3UW7 scheme. 
}
  \label{fig:zstep_contours}
\end{figure}

Besides much better energy conservation properties, we find the TVDRK3UW7 scheme to be very 
robust against spurious Gibbs oscillations, even though it is not a shock-capturing scheme.
This is clearly visible in~\fig{fig:zstep_contours}, where we plot the time history of the 
profiles of $\Apar$ and $n_e$ obtained with both schemes for $\hat k=1$. 
As can be seen, the TVDRK3UW7 scheme advects the initial condition with no visible deterioration, 
unlike the MacCormack scheme. 

\subsection{Linear Kinetic Alfv\'en Wave}
\label{sec:KAW}

The linearisation of~\eqs{eq:mom}{eq:ge} in the collisionless limit yields the kinetic Alfv\'en wave dispersion relation~\cite{zocco_reduced_2011}:

\begin{equation}
\label{eq:LKAW}
\[\zeta^2-\frac{\tau}{Z}\frac{\kperp^2d_e^2/2}{1-\Gamma_0(\kperp^2\rhoi^2/2)}\]
\[1+\zeta Z(\zeta)\]=
\frac{1}{2}\kperp^2d_e^2,
\end{equation}
where $\zeta = \omega / |\kpar| v_{the}$, $Z(\zeta)$ is the plasma dispersion function and $\kperp^2=k_x^2+k_y^2$. 

\begin{figure}
  \center
  \includegraphics[width=0.49\textwidth]{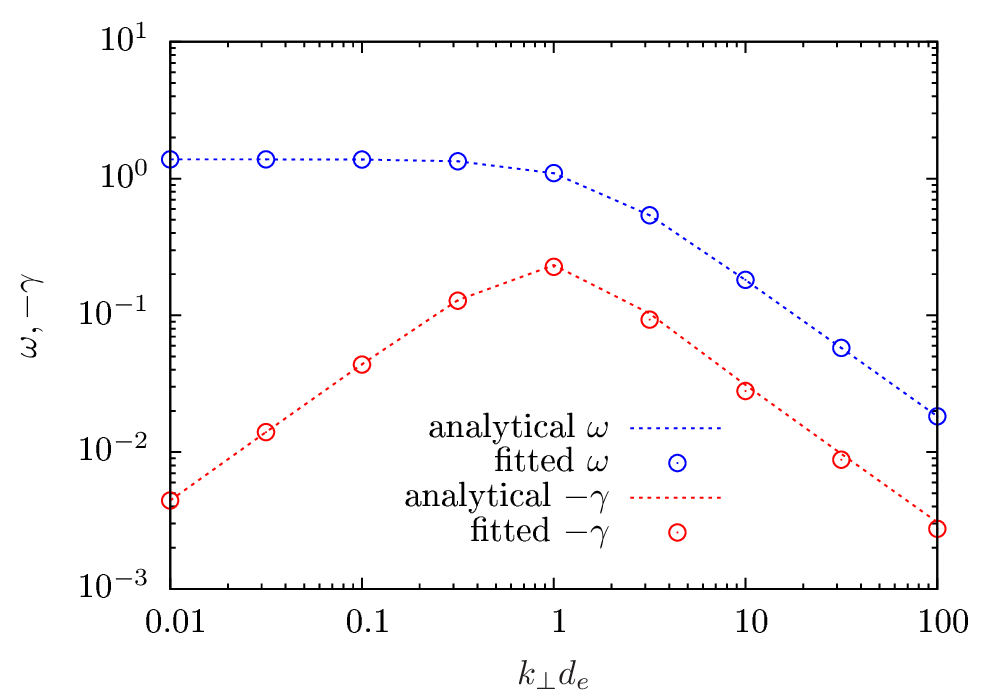}
  \includegraphics[width=0.49\textwidth]{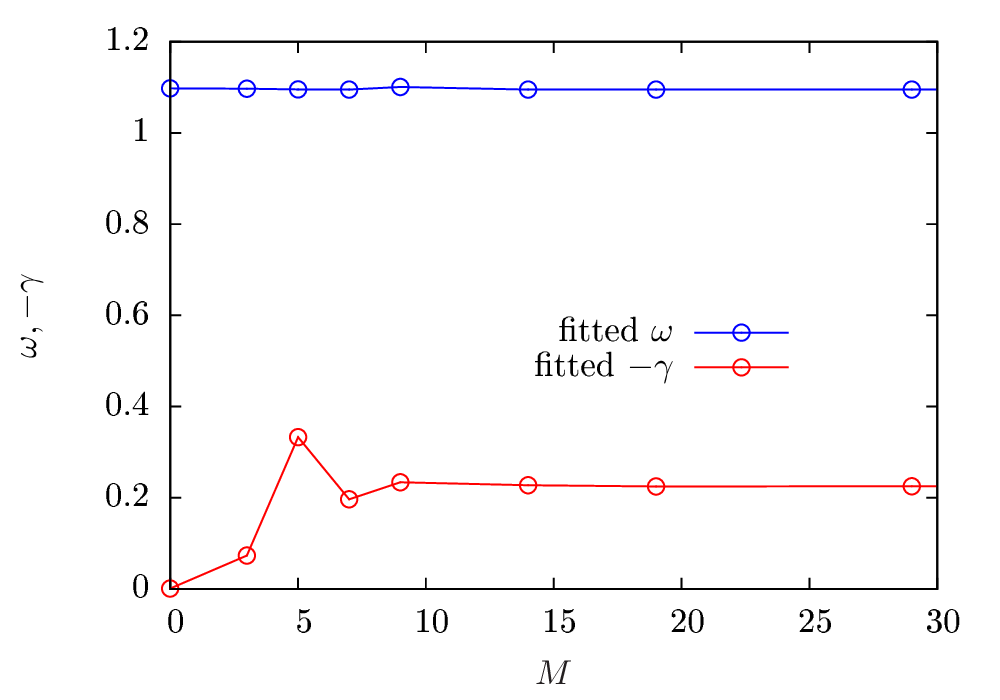}
  \caption{Left: Frequency and damping rate of the kinetic Alfv\'en wave (KAW), in units of $\tau_A$, at fixed 
$\kperp \rho_i=1$, $\tau=1$, $M=19$, as a function of the electron skin-depth 
$d_e=c/\omega_{pe}$. Lines are the exact solution of the analytical dispersion relation,~\eq{eq:LKAW}, 
whereas data points are obtained from \viriato. Right: KAW frequency and damping rate obtained from \viriato~for 
fixed $\kperp d_e=1$, as a function of the total number of Hermite moments kept, $M$.
}
  \label{fig:KAW_DR}
\end{figure}

On the left plot of~\fig{fig:KAW_DR} we show a comparison between the analytical values of the frequencies and damping rates, obtained by 
solving~\eq{eq:LKAW}, and 
those computed by \viriato~setting the number of Hermite moments to $M=19$ and the number of grid points in the $z$-direction to $32$. 
Very good agreement is observed 
over several orders of magnitude of the electron 
skin depth, $d_e$; the maximum relative error, obtained for the highest value of $d_e$, is only a few percent.
The right plot shows the values of the frequency and damping rate for $\kperp d_e=1$ as a function of the number of Hermite moments.  
For $M\ge 9$ the damping rate converges to the analytical value ($-\gamma=0.2331$), whereas for $\omega$ very little dependence on $M$ is observed.

\subsection{Tearing Mode}
\label{sec:tearing}
The tearing mode~\cite{furth_finiteresistivity_1963} is a fundamental plasma 
instability driven by a background current gradient.  
Tearing leads to the 
opening, growth and saturation of (one or more) magnetic island(s) via the reconnection of a background 
magnetic field. 
It is of intrinsic interest to magnetic confinement fusion devices,
where it occurs either in standard or modified form (i.e., neoclassical tearing, microtearing).
It also represents the most basic paradigm for studies of magnetic reconnection.

In this section, we present the results of a linear benchmark
of \viriato~against the gyrokinetic code {\tt AstroGK}~\cite{numata_astrogk:_2010} for the tearing mode problem. 
We consider an
in-plane magnetic equilibrium configuration given by $B_{y,eq}=-dA_{\parallel,eq}/dx$, with
$A_{\parallel,eq}=A_{\parallel 0}/\cosh^{2}(x/a)$, with $a$ the normalizing equilibrium scale length.
The simulations are performed in a doubly periodic box of dimensions $L_x\times L_y$, with 
$L_x/a=2\pi$ and $L_y=2.5\pi$, such that $\hat k_y= 2\pi a/L_y$ yields  
the tearing instability parameter 
$\Delta' a =2(5-\hat k_y^2)(3+\hat k_y^2)/(\hat k_y^2\sqrt{4+\hat k_y^2})\approx 23$.
Other parameters are $\rho_i/a=0.2$, $\tau=1$, $d_e/a=0.037$. All \viriato~simulations keep $M=10$.

\begin{figure}
  \center
  \includegraphics[width=8cm]{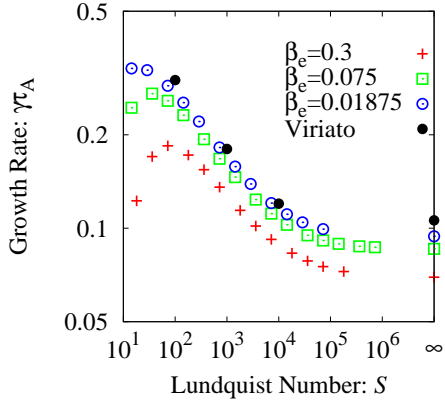}
  \caption{Tearing mode growth rate as a function of the Lundquist number.
Figure shows a comparison between the results obtained with the gyrokinetic code 
{\tt AstroGK}~\cite{numata_astrogk:_2010} for varying values of $\beta_e$ and \viriato. 
As expected, good agreement is obtained in the small $\beta_e$ limit.}
  \label{fig:tear_bench_linear}
\end{figure}

\fig{fig:tear_bench_linear} shows a plot 
of the linear growth rate of the tearing mode as a function of the Lundquist number
$S=a v_A/\eta$. The $S=\infty$ case is obtained by setting $\eta=0$, in which case the tearing mode is collisionless, i.e., the frozen-flux condition is broken by electron inertia instead. 
Calculations with
{\tt AstroGK}  are done at three different values 
of $\beta_e$ and mass ratio:
$(\beta_e,m_e/m_i)=(0.3,0.01),~(0.075,0.0025),~(0.01875,6.25\times10^{-4})$ 
(crosses, squares and circles, respectively; this is the same data as plotted in Fig. 2 of 
Ref.~\cite{numata_gyrokinetic_2011}). 
As seen, the agreement between the two codes improves for smaller $\beta_e$, and is rather 
good for the smallest value of $\beta_e=0.01875$. Though it is expected that 
gyrokinetics will converge to KREHM as $\beta_e$ is decreased, we note that, at least  
in this particular case, agreement is achieved for $\beta_e$ substantially larger than $m_e/m_i$ (a 
factor of $30$),
suggesting that KREHM may remain a reasonable approximation to the plasma dynamics outside its
strict asymptotic limit of validity set by the requirement $\beta_e\sim m_e/m_i$.

A nonlinear benchmark is provided by 
the comparison of the tearing mode saturation amplitude with
the prediction of MHD theory~\cite{militello_simple_2004, escande_simple_2004,
loureiro_x-point_2005}. This was reported in Ref.~\cite{loureiro_fast_2013}, where it is shown
that \viriato~accurately reproduces the theoretical prediction in the parameter region 
where such prediction is valid [i.e., for $\Delta'a\sim1$ and as long as islands are larger than the 
kinetic scales of the problem ($\rho_i,~\rho_s,~d_e$)].

Finally, see also Figs. 1 and 3 of Ref.~\cite{numata_ion_2014} for more direct comparisons between \viriato~and {\tt AstroGK} in the linear and nonlinear regime of a collisionless tearing mode simulation.

\subsection{Orszag-Tang vortex problem}
\label{sec:OT}
The Orszag-Tang (OT) vortex problem~\cite{orszag_small-scale_1979} 
is a standard nonlinear test for fluid codes, and a basic paradigm
in investigations of decaying MHD 
turbulence~\cite{orszag_small-scale_1979,biskamp_dynamics_1989,politano_1995,biskamp_two-dimensional_2001}.
Here we present results from a series of 2D and 3D runs, including 
a kinetic case. For easy reference, we summarise the main parameters of each simulation 
performed in \tabref{OT_table}.\\
\begin{table}
  \centering
  \begin{tabular}{cccccc}
    Run & Dim. & \#Gridpoints &$\rho_i/a$  & Dealiasing & Hyper-diss.?\\
    \hline
    \hline
    A & 2D & $2048^2$ & 0 & $2/3$'s rule & no\\
    A1 & 2D & $2048^2$ &  0 & $2/3$'s rule & yes\\
    B & 2D & $2048^2$ & 0  & Hou-Li & no\\
    B1 & 2D & $2048^2$ & 0  & Hou-Li & yes\\
    \hline
    C & 3D & $512^3$ & 0  & Hou-Li & yes\\
    D & 3D & $256^3$ & 2  & Hou-Li & yes\\
    E & 3+1D & $256^3$ & 2  & Hou-Li & yes\\
    \hline
    \hline
  \end{tabular}\\
  \caption{Main parameters for decaying turbulence runs [with the Orszag-Tang-type 
    initial conditions of \eqs{modOT-phi}{modOT-psi} for the 2D runs, and of 
    \eqs{modOT3D-phi}{modOT3D-psi} for the 3D runs]. In all cases, $\rho_s=\rho_i$ and $d_e=0$. Run E also includes $20$ Hermite moments.}
    \label{OT_table}
\end{table}

\subsubsection{2D simulations of the OT vortex problem}
\label{sec:2D_OT}
To avoid an overly symmetric initial configuration, we adopt the 
modification of OT initial conditions proposed in Ref.~\cite{biskamp_dynamics_1989},
namely\footnote{We note 
for completeness that we have also performed a simulation with the same (symmetric) 
initial condition as used in Ref.~\cite{numata_astrogk:_2010} and
 obtained excellent agreement with the results reported there.}:
\begin{align}
\label{modOT-phi}
\Phi(x,y) = \cos(2\pi x/L_x + 1.4) + \cos(2\pi y/L_y+0.5),\\
\label{modOT-psi}
\Psi(x,y) = \cos(4\pi x/L_x+2.3) + \cos(2\pi y/L_y + 4.1).
\end{align}
The runs are performed on a box of dimensions $L_x=L_y=2\pi$, at a resolution of 
$N_x\times N_y=2048^2$ collocation points. In the cases where no 
hyper-dissipation is used (runs A and B), the resistivity is set to $\eta=10^{-3}$, 
and the magnetic Prandtl number $Pm=\nu/\eta=1$. 
The kinetic scales $\rho_i, ~\rho_s,~d_e$ are set to zero, so this is strictly a RMHD 
run.

\begin{figure}
  \center
  \includegraphics[width=\textwidth]{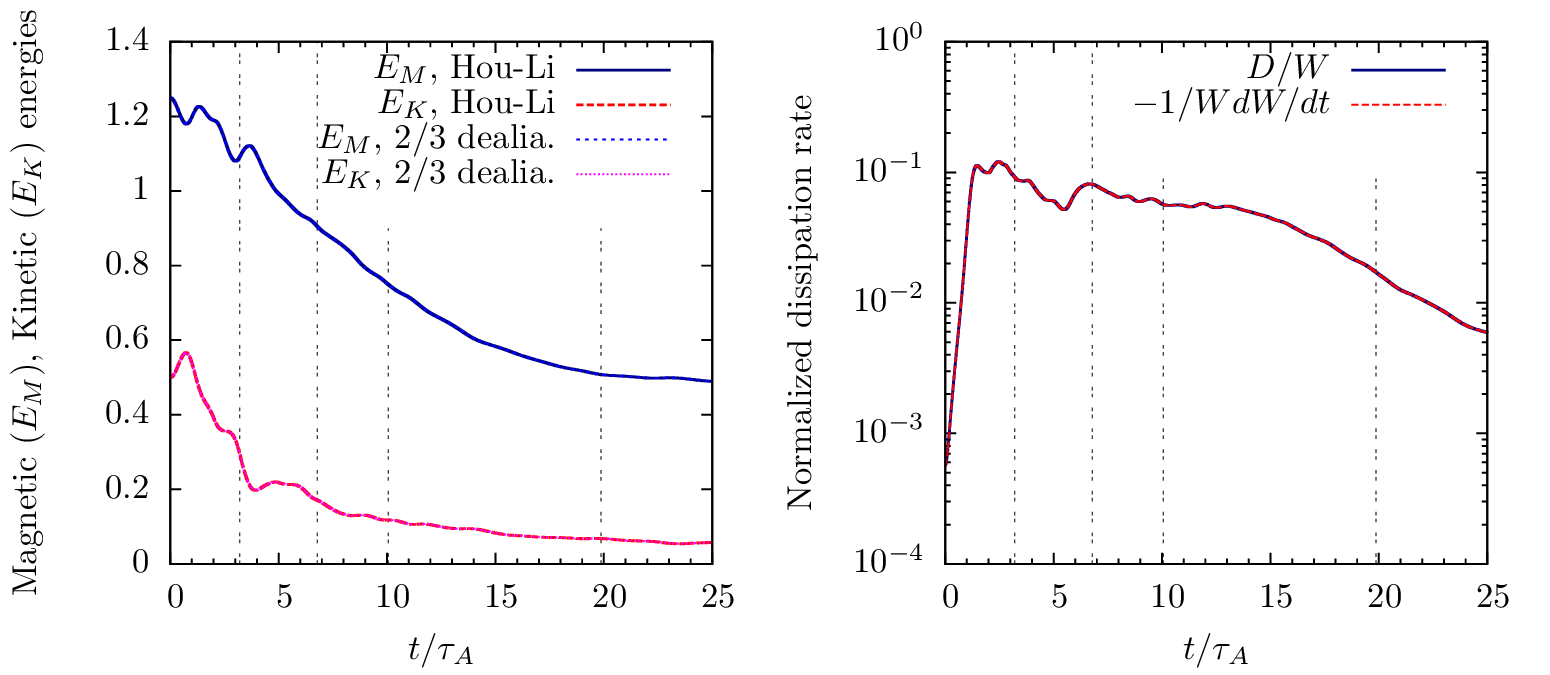}
  \caption{Runs A and B. 
    Left panel: Time traces of the magnetic ($E_M$) and kinetic ($E_k$) energies, 
    obtained from runs with different dealiasing methods: ``Hou-Li'' uses the 
    high-order Fourier filter of Ref.~\cite{hou_computing_2007}, given by \eq{hou_li_filter}; 
    ``$2/3$ dealia.'' uses the 
    usual $2/3$'s rule of Ref.~\cite{orszag_elimination_1971}, \eq{two_thirds_rule}.
    Right panel: Time trace of the energy dissipation rate (for the Hou-Li run), 
    normalized by the instantaneous 
    total energy, $D/W$. 
    Overplotted is $-1/W dW/dt$: code conserves energy to better than $0.1\%$
    in this run. The vertical dotted lines identify the times at which the contours 
    of \fig{fig:OTBW_2D_pannel} and spectra of 
    \fig{fig:OTBW_2D_spectra} are plotted.}
  \label{fig:OTBW_2D_traces}
\end{figure}

Magnetic ($E_M$) and kinetic ($E_K$) energy time traces for runs A and B are shown on the 
left-hand panel of \fig{fig:OTBW_2D_traces}. We compare the results obtained 
using the Hou-Li high order Fourier filter, \eq{hou_li_filter}, with 
those obtained with the standard $2/3$'s dealiasing rule of 
Orszag~\cite{orszag_elimination_1971}, \eq{two_thirds_rule}. 
The agreement between the two sets of results is perfect, demonstrating that 
the Hou-Li filter does as good a job at conserving energy as the $2/3$'s rule.

The right-hand panel shows the time trace of the energy dissipation, normalized 
by the instantaneous total energy, i.e., 
\be
\frac{D}{W} \equiv \frac{\int dV (\eta j^2 + \nu \omega^2)}{\frac{1}{2}\int dV (B^2 + u^2)}.
\ee
Since no energy is being injected into the system, the RMHD equations should obey the conservation relation 
\be
\label{RMHD_inv}
\frac{dW}{dt}=-D.
\ee
In order to demonstrate the accuracy of the code, we overplot a time trace of 
$-1/W dW/dt$. The very good agreement between the two curves is manifest; 
in this particular run, \eq{RMHD_inv} is satisfied to better than $0.1\%$.

\begin{figure}
  \center
  \includegraphics[width=\textwidth]{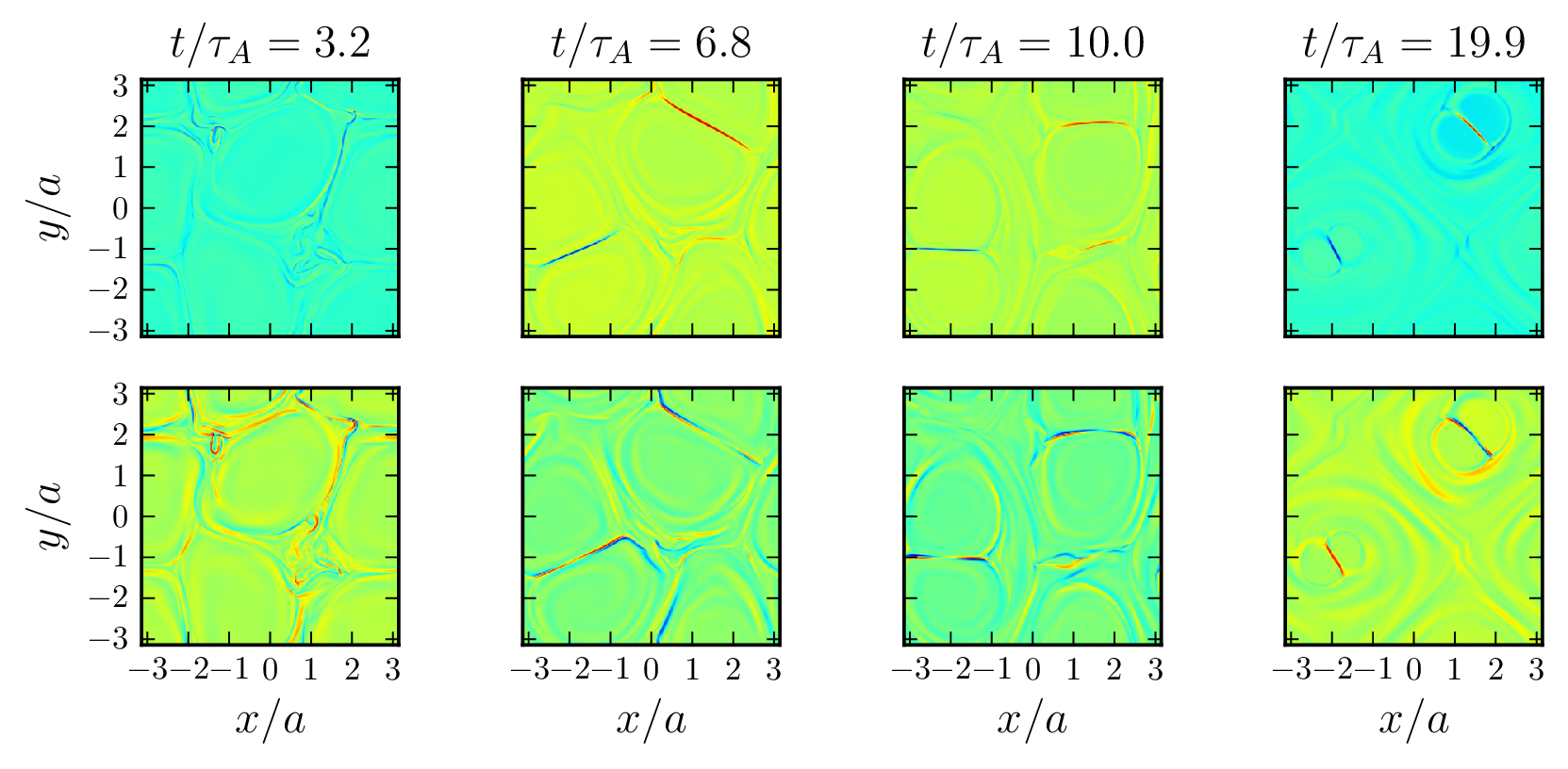}
  \caption{Run B (2D). Contour plots of current (top row) and vorticity (bottom row) 
at different times (identified by the vertical lines in \fig{fig:OTBW_2D_traces}).}
  \label{fig:OTBW_2D_pannel}
\end{figure}

Contour plots of current and vorticity (i.e., $\lapperp \Phi$) at the times
identified by the vertical lines in \fig{fig:OTBW_2D_traces} are plotted in \fig{fig:OTBW_2D_pannel} (top and bottom rows, respectively).
The formation of sharp current and vorticy sheets is observed, as expected. At $t/\tau_A=10.0$ one can observe a plasmoid~\cite{loureiro_instability_2007,loureiro_plasmoid_2013} erupting from the current sheet on the lower right-hand corner of the plot, in what is perhaps the small-scale version of the observations reported in Ref.~\cite{loureiro_turbulent_2009}. 
The role of the tearing instability of current sheets in 2D decaying turbulence has been previously discussed in Refs.~\cite{biskamp_dynamics_1989, biskamp_two-dimensional_2001}.

\begin{figure}
  \center
  \includegraphics[width=0.7\textwidth]{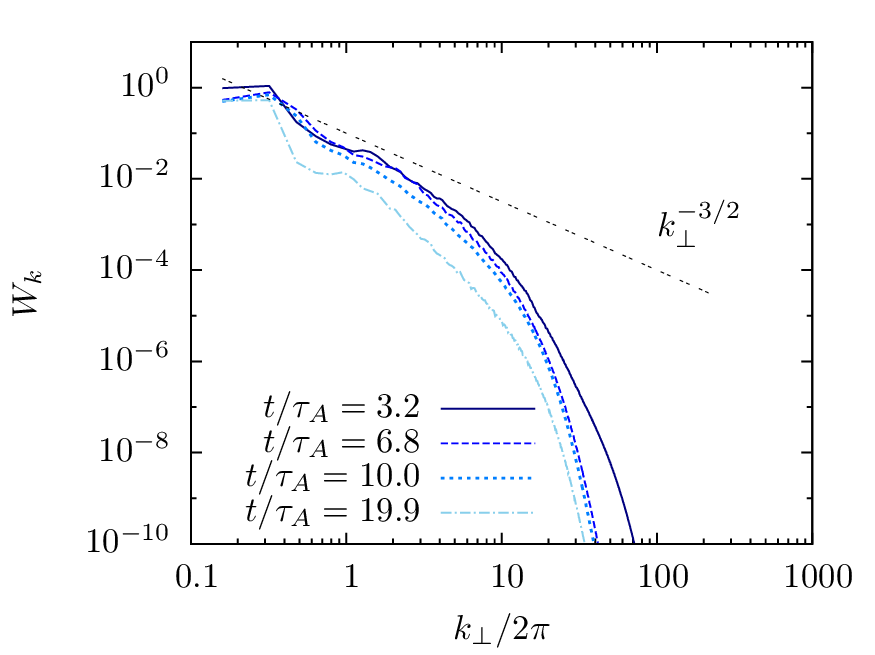}
  \caption{Run B (2D). Total energy spectra at different times (identified by 
the vertical lines in \fig{fig:OTBW_2D_traces}). A $k_\perp^{-3/2}$ slope 
is shown for reference.}
  \label{fig:OTBW_2D_spectra}
\end{figure}

\fig{fig:OTBW_2D_spectra} shows the total energy spectra obtained from 
the simulation with the Hou-Li filter (run B), taken at the times 
identified by the vertical lines in \fig{fig:OTBW_2D_traces}.
There is no evidence of pile-up (bottleneck) at the small scales (we note that 
the only dissipation terms present in this simulation are the standard laplacian 
resistivity and viscosity, i.e., there is no hyper-dissipation).
Due to the relatively large values of the dissipation coefficients used in this simulation, the inertial range is very limited and it is not possible to clearly fit a unique power law; 
for reference, $\kperp^{-3/2}$ is indicated in~\fig{fig:OTBW_2D_spectra}, following the Iroshnikov-Kraichnan prediction~\cite{iroshnikov_turbulence_1963, kraichnan_inertial_1965}, and its numerical confirmation reported in Refs.~\cite{biskamp_dynamics_1989, biskamp_two-dimensional_2001}
(although steeper power-laws $\sim\kperp^{-5/2}$ have also been reported in the
literature~\cite{kinney_coherent_1995,dellar_lattice_2013}).

\begin{figure}
  \center
  \includegraphics[width=0.7\textwidth]{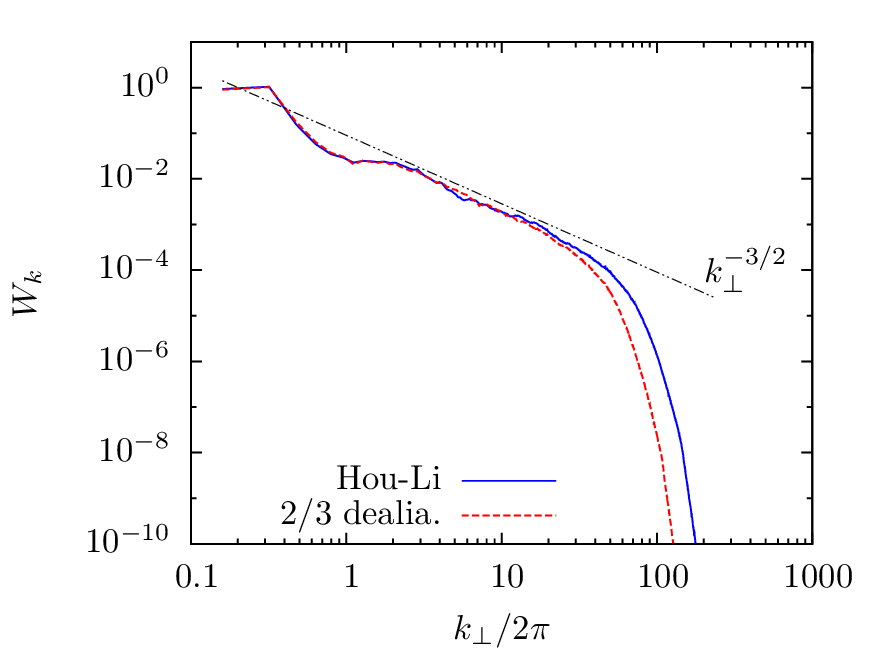}
  \caption{Runs A1 and B1 (2D). Total energy spectra at $t/\tau_A\approx 6.0$ 
    obtained with the Hou-Li filter (blue, full line) and 
    with the standard $2/3$'s dealising rule (red, dashed line). The Hou-Li method results 
    in an extended inertial range for the same number of collocation points, 
    as expected. Neither spectra shows signs of energy pile-up at the small scales.
    The power-law $k_\perp^{-3/2}$ is indicated for reference.}
  \label{fig:OTBW_2D_spectra_hyper}
\end{figure}

A much longer and cleaner inertial range is obtained by replacing the standard (laplacian) 
dissipation terms with hyper-dissipation (runs A1 and B1). In that case, the spectra shown in 
\fig{fig:OTBW_2D_spectra_hyper} are obtained; the inertial range now shows an excellent agreement with the power-law slope of $-3/2$. Note also the extended inertial range obtained when the Hou-Li filter is used (B1) instead of the standard $2/3$'s dealiasing.

\subsubsection{3D simulations of the OT vortex problem}
\label{sec:3D_OT}
For the 3D simulations the initial conditions differ from the 2D case only in that they are 
modulated in the $z$-direction, as follows:
\begin{align}
\label{modOT3D-phi}
\Phi(x,y) = \[\cos(2\pi x/L_x + 1.4) + \cos(2\pi y/L_y+0.5)\]\sin(2\pi z/L_z),\\
\label{modOT3D-psi}
\Psi(x,y) = \[\cos(4\pi x/L_x+2.3) + \cos(2\pi y/L_y + 4.1)\]\cos(2\pi z/L_z).
\end{align}

We perform three different runs with these
initial conditions (runs C, D and E). 
The first (run C) is just a straightforward 
extension to 3D of run B1, except now 
with a resolution of $N_x \times N_y\times N_z = 512^3$.
The second (run D) is designed to look at sub-ion-Larmor radius turbulence (i.e., kinetic Alfv\'en wave turbulence); thus we set
$\rho_i/a = 2 , ~d_e/a = 0.01 $, where $a=L_x/(2\pi)$, and $\tau=1$. The 
resolution in this case is $N_x\times N_y \times N_z = 256^3$ 
(we use a smaller resolution here because the timestep, which is set
by the CFL condition, is now also smaller, due to the dispersive 
nature of the kinetic Alfv\'en waves). 
Finally, run E also includes the velocity-space dependence, represented with $20$ 
Hermite moments (meaning that it differs from run D in that the electrons are no longer isothermal, i.e., $g_e\ne0$)

\begin{figure}
  \center
  \includegraphics[width=0.7\textwidth]{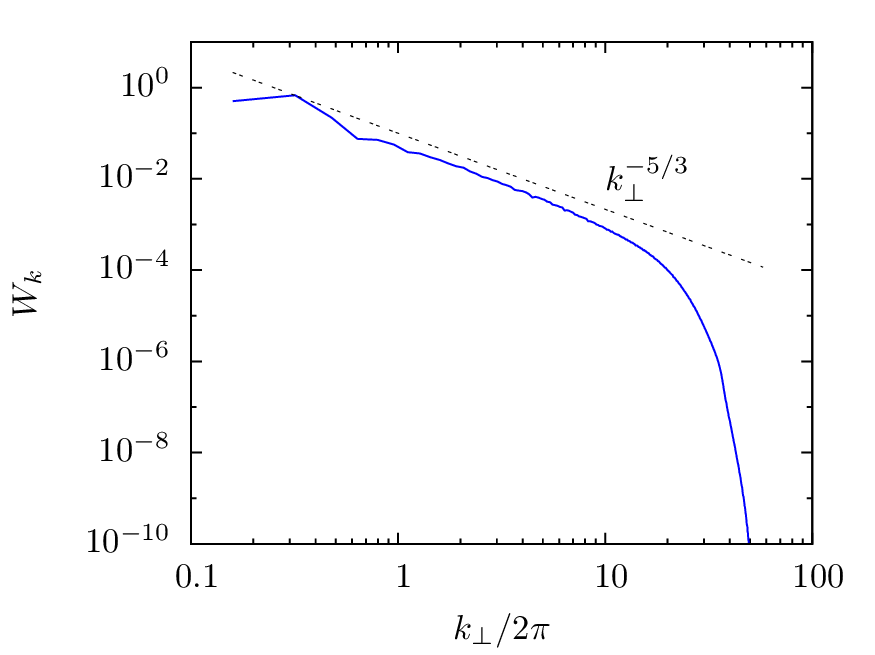}
  \caption{Run C (3D). Total energy spectra at $t/\tau_A\approx 4.0$. 
    A $k_\perp^{-5/3}$ slope is shown for reference.}
  \label{fig:OTBW_3D_spectra}
\end{figure}

The total energy spectrum obtained for run C is shown in \fig{fig:OTBW_3D_spectra}.
The inertial range shows very good agreement with the Goldreich-Sridhar $k^{-5/3}$ power law~\cite{goldreich_toward_1995} and again is clean of bottleneck effects.

\begin{figure}
  \center
  \includegraphics[width=0.7\textwidth]{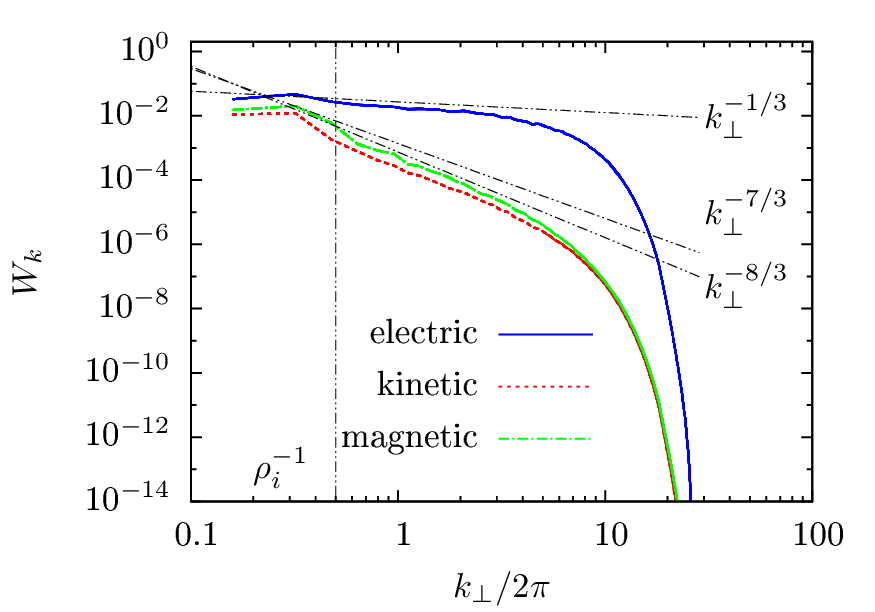}
  \caption{Run D (3D). Spectra for decaying turbulence [with OT-like 
initial conditions, \eqs{modOT3D-phi}{modOT3D-psi}] at $t/\tau_A\approx 2.2$. 
The blue (full) line represents the perpendicular electric energy spectrum;
the red (dashed) line is the perpendicular 
magnetic field energy and the green (dash-dot) line is the kinetic energy. 
The slopes $\kperp^{-1/3}$, $\kperp^{-7/3}$ and $\kperp^{-8/3}$ are 
indicated for reference (see text for discussion). 
The vertical line indicates the ion Larmor radius scale.}
  \label{fig:OTBW_3D_KAW_spectra}
\end{figure}

\fig{fig:OTBW_3D_KAW_spectra}
shows the magnetic, kinetic and electric energy spectra for run D, where we are now focussing on 
sub-ion Larmor radius scales.
The slopes indicated refer to several power laws that have been widely discussed in the
literature. 
In particular, we see that the separation between electric and magnetic energy scalings, 
occurring at around $(k_\perp/2\pi)\rho_i\sim1$, agrees 
quite well with the solar wind observations reported by Bale {\it et al.}~\cite{bale_2005} and with the gyrokinetic simulations of Howes \etal~\cite{howes_kinetic_2008}.
However, instead of the $-7/3$ power law for the magnetic energy 
suggested in those works (discussed in more detail in Ref.~\cite{schekochihin_astrophysical_2009}), 
we see that our data seems to more closely fit a $-8/3$ scaling, which is a better fit to the $-2.8$ slope often reported in observations (e.g., \cite{alexandrova_universality_2009}) and in agreement with 
the recent work of Boldyrev and Perez~\cite{boldyrev-perez-2012} 
on strong kinetic Alfv\'enic turbulence.

\begin{figure}
  \center
  \includegraphics[width=0.7\textwidth]{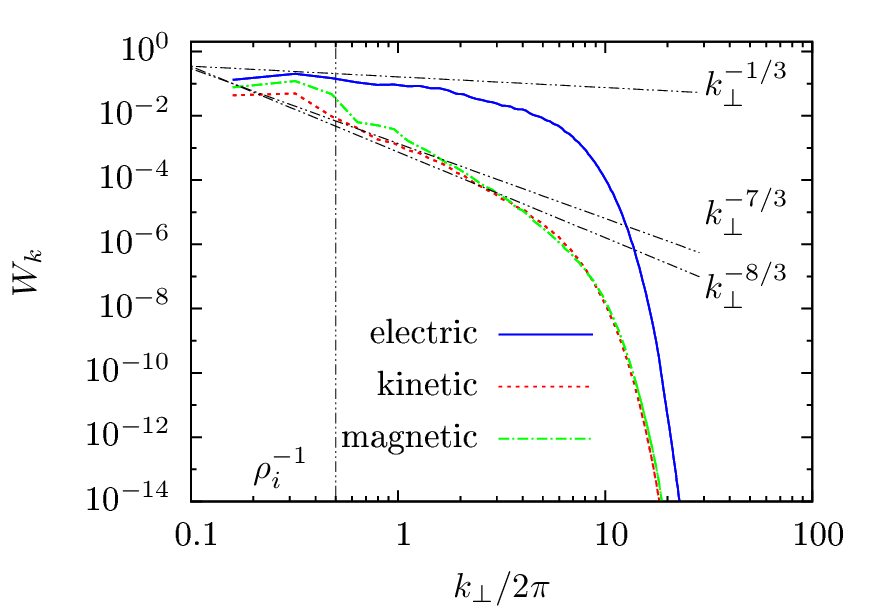}
  \caption{Run E (3D, with $20$ Hermite moments). Spectra at $t/\tau_A\approx 2.2$ for OT-decaying kinetic turbulence. Lines represent
the same quantities as in~\fig{fig:OTBW_3D_KAW_spectra}. 
See text for a discussion of the power laws indicated.}
  \label{fig:OTBW_3D_KAW_spectra_node}
\end{figure}

\begin{figure}
  \center
  \includegraphics[width=0.7\textwidth]{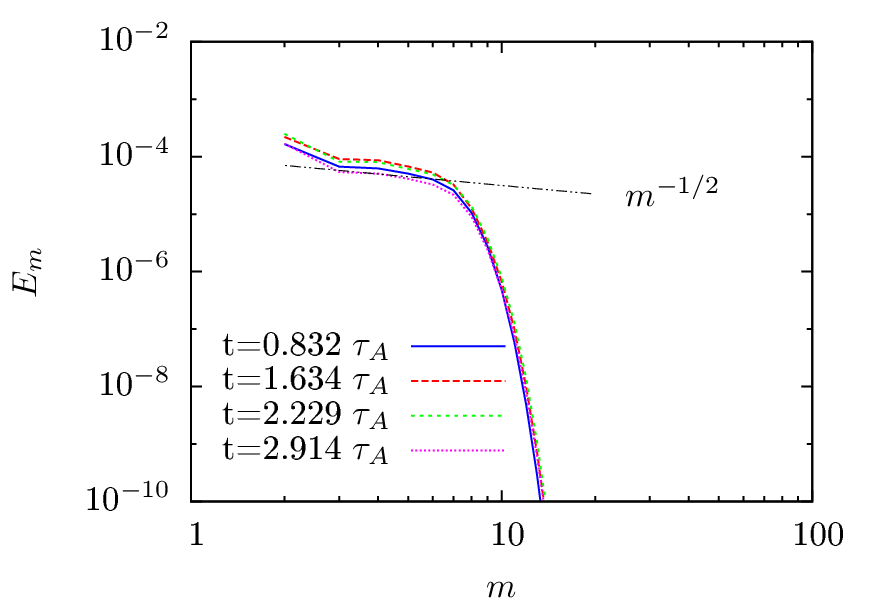}
  \caption{Run E (3D, with $20$ Hermite moments). Electron free-energy spectra $E_m=|g_m^2/2|$ at different times. An indicative power law of $m^{-1/2}$ for the inertial range
is also shown~\cite{zocco_reduced_2011}.}
  \label{fig:OTBW_3D_KAW_gm_spectra}
\end{figure}

\fig{fig:OTBW_3D_KAW_spectra_node} again shows energy spectra, this time for run E, which differs from run D in that it also includes Hermite 
moments (i.e., it is a fully kinetic run, whereas D assumes isothermal electrons, $g_e=0$).
Comparing the magnetic spectra in the two cases (i.e, runs D and E, both drawn at the same time), we see that 
its values increase at the larger (spatial) scales when adding the Hermite moments, by about an order of magnitude, and run E's spectrum seems to be somewhat steeper than $-8/3$.
Such differences may be due to Landau damping, which is present in run E, but absent in run D. 
The Hermite spectrum (i.e., the electron free energy spectrum, $E_m=|g_m^2|/2$) for run E is shown 
in~\fig{fig:OTBW_3D_KAW_gm_spectra}, at different times. 
A $-1/2$ slope is indicated for reference; this is the inertial-range slope predicted by 
Zocco \& Schekochihin~\cite{zocco_reduced_2011} for the linear phase-mixing of Kinetic Alfv\'en waves. Since the number of Hermite moments ($20$) used is quite small we get an equivalently  
limited inertial range, and thus the agreement with the $-1/2$ slope can only be regarded as indicative; however, this tentative agreement lends credence to the idea that Landau damping may be playing a significant role in this simulation.  
A detailed analysis of kinetic turbulence in the KREHM framework and, in particular, of the relative importance of the different energy dissipation mechanisms available, will be the subject of a future publication.

\begin{figure}
  \center
  \includegraphics[width=0.49\textwidth]{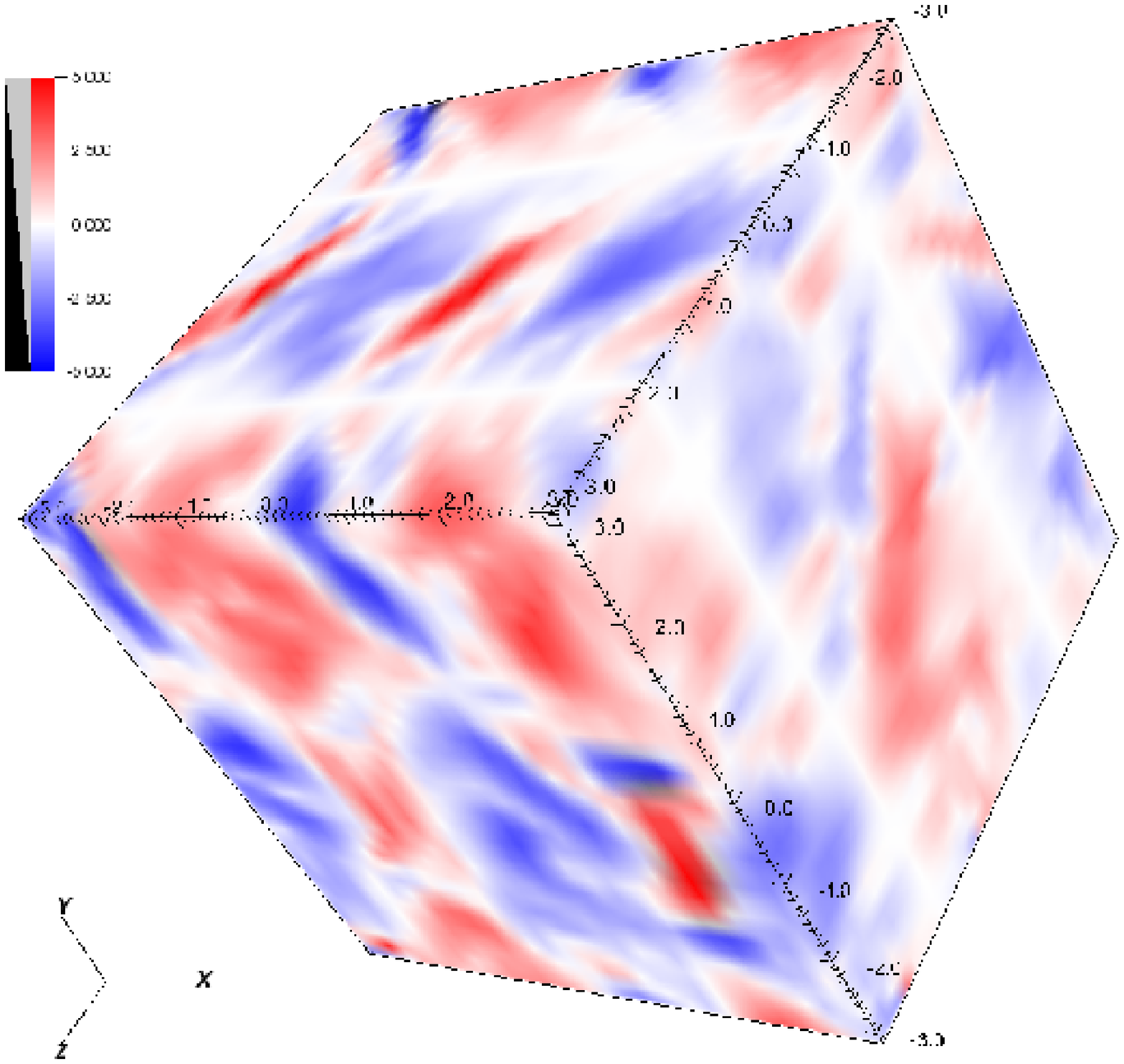}
  \includegraphics[width=0.49\textwidth]{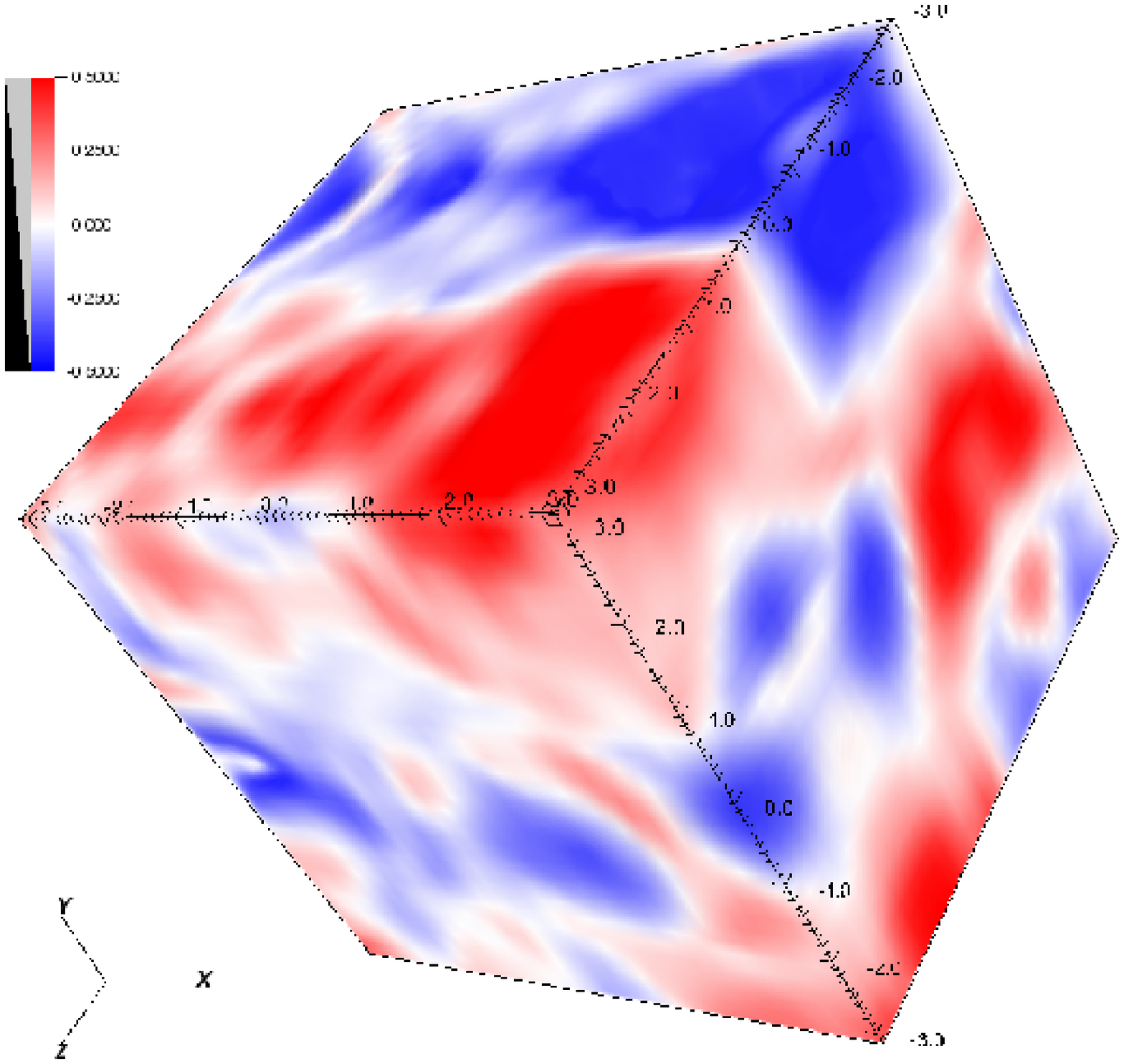}
  \caption{Run E: contour plots of the parallel electron velocity, $\ue$ (left), and density perturbations $n_e$ (right), at $t/\tau_A\approx 2.2$.} 
  \label{fig:Apar_t_2.2}
\end{figure}

Finally, for completeness we show in \fig{fig:Apar_t_2.2} contour plots of the electron parallel velocity, $\ue$, and of the density perturbations, $n_e$, taken at the same time as the spectra of
\fig{fig:OTBW_3D_KAW_spectra_node} ($t/\tau_A\approx 2.2$). 

\subsection{Collisionless damping of slow modes}
\label{sec:slow_mode_tests}
We turn now to a benchmark of \viriato's implementation of the KRMHD equations.
Linearly, slow modes in KRMHD are subject to collisionless damping via the Barnes damping
mechanism~\cite{barnes66}. An initial perturbation damps at a rate that depends on the
parameter $\Lambda$. If slow mode fluctuations are constantly driven with an external
force (this is achieved by adding a forcing term to \eqref{eq:slowmodekin}), then the
system can be thought of as a plasma-kinetic Langevin equation. The mean-squared amplitude
of the electrostatic potential for such a system reaches a steady-state saturation level,
which can be derived analytically~\cite{kanekar_fdr_2014}.

\begin{figure}
    \center
    \includegraphics[width=9cm]{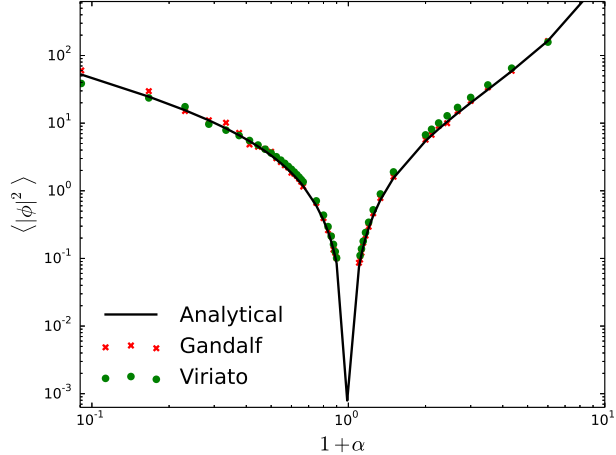}
    \caption{Steady-state amplitude of the electrostatic potential vs $1+\alpha$, where
    $\alpha = -1/\Lambda$. The solid line is the analytical
    prediction~\cite{kanekar_fdr_2014}, the red crosses are numerical results calculated
    using {\tt Gandalf}, and the green circles are calculated using \viriato.}
    \label{fig:slowmode}
\end{figure}

In \figref{fig:slowmode}, we compare the steady-state saturation levels computed using
\viriato~with the analytical predictions, and the numerical results from another code --- {\tt Gandalf} 
(a fully spectral GPU code that solves the KRMHD equations). Slow mode fluctuations
were driven using white noise forcing\footnote{Another way of forcing the system which is also implemented 
in \viriato~is via an oscillating Langevin antenna~\cite{tenbarge_antenna_2014}.}
which injected energy into the system with unit
power. The spatial resolution was set to $N_x\times N_y \times N_z = 32^3$; $20$ Hermite
moments of the distribution function were retained, $M=20$. The system was evolved
until it reached a steady state. The saturation level was then calculated by averaging over the
steady state fluctuations for a few Alfv\'en times. It can be seen that the saturation amplitudes obtained
using \viriato~are in near perfect agreement with those calculated by {\tt Gandalf}, as well as with the analytical prediction.

\section{Performance}
\label{sec:perform}

\viriato~has been used on a variety of computing clusters, with different architectures. 
It is quite easy to install and run, having dependencies only on standard, widely-used 
libraries such as LAPACK~\cite{lapack_1999} and FFTW~\cite{FFTW05}. 
Its parallelization relies on standard MPI routines.
 
As described in detail in~\Secref{sec:numerics}, the direction parallel to the field can be integrated 
by two different numerical methods, both of them fairly scalable, in terms of parallel performance. 
In contrast, the direction perpendicular to ${\bf B}_0$ 
uses standard pseudospectral techniques, which are plagued with well-known limits on scalability, due to 
the inherent non-locality of Fourier transforms. 
For this reason, if one wishes to increase the number of processors 
for a given computation, it is more effective to do so by increasing the 
ratio between the number of processes for the parallel direction and the
number of processes in the perpendicular direction. 

\begin{figure}
  \center
  \includegraphics[width=8cm]{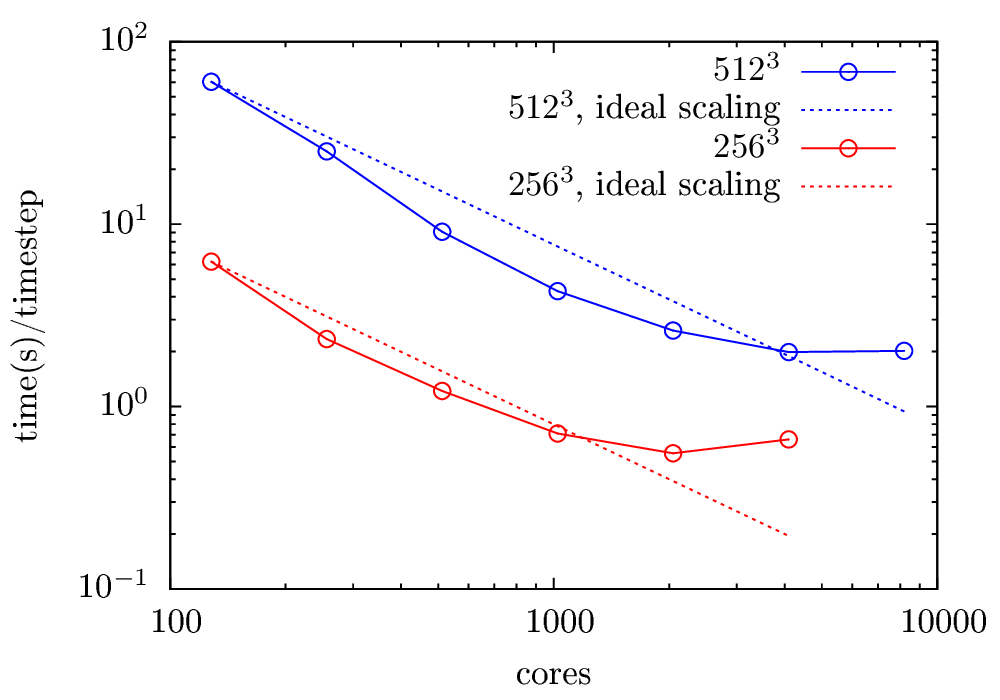}
  \caption{\viriato~timings measured on the Helios cluster, for two different fixed problem 
sizes (strong scaling). 
A supralinear trend can be observed, which breaks down after $1024$ cores for the $256^3$ case and 
at $\sim4096$ cores for the $512^3$ one. The vertical axis gives the wall-clock time (in seconds) 
spent per timestep.
}
  \label{fig:helios_times}
\end{figure}

The results of such a test, made on the Helios machine (an Intel Xeon E5 cluster), 
can be seen on~\fig{fig:helios_times}, where the 
MacCormack method was used in the parallel 
direction. The initial conditions are the 3D Orszag-Tang vortex given by~\eqs{modOT3D-phi}{modOT3D-psi}, with 
$15$ Hermite moments. 
We look at strong scaling, keeping the problem size fixed and varying the number of 
MPI processes, mainly in the parallel direction. 
This produces a supralinear scaling, which breaks down 
after $1024$ cores for the $256^3$ case and at $\sim4096$ cores for the $512^3$ one. 
Similar results have been obtained 
on other clusters, such as Stampede (a mixed Intel Xeon E5 and Intel Xeon Phi Coprocessor cluster), 
Hopper (a Cray XE6) and Edison (a Cray XC30.

Currently ongoing optimization work includes parallelizing 
the computation of the Hermite moments' via OpenMP.
  
\section{Conclusions}
\label{sec:summary}
This paper describes \viriato, a novel code developed to investigate strongly magnetised, 
weakly-collisional, fluid-kinetic plasma dynamics in (2D or 3D) slab geometry. \viriato~solves two different sets of equations: 
the Kinetic Reduced Electron Heating Model (KREHM) of Zocco \& Schekochihin~\cite{zocco_reduced_2011} (which simplifies to conventional reduced-MHD~\cite{kadomtsev_nonlinear_1974, 
strauss_nonlinear_1976} in the appropriate limit)
and the Kinetic Reduced MHD (KRMHD) equations of 
Schekochihin \etal~\cite{schekochihin_astrophysical_2009}.

The main numerical methods and the overall algorithm are described. 
A noteworthy feature of \viriato~is its spectral representation of velocity-space, achieved via a Hermite expansion
of the distribution function, as proposed in~\cite{zocco_reduced_2011} for KREHM and in~\cite{kanekar_fdr_2014} for the KRMHD equations. This representation has the attractive property of converting the kinetic 
equation for the distribution function into a coupled set of fluid-like equations for each Hermite polynomial coefficient --- the advantage being that such equations are numerically more convenient to solve than the kinetic equation where they stem from. On the other hand, the Hermite expansion introduces a closure problem (in the sense that the equation for the Hermite coefficient of order $m$ couples to that of order $m+1$). To address this problem, we present a nonlinear, asymptotically rigorous closure whose validity requires only that collisions are finite, but otherwise as small as required. Naturally, the smaller the collision frequency the higher the number of Hermite moments that need to be kept to guarantee the accuracy of the closure. 
Realistic values of the collision frequency in the systems that are of primary interest to us (e.g., modern fusion devices, space and astrophysical environments) lead to impractically large number of moments. 
The adoption of a hyper-collision operator (the direct translation into Hermite space of the usual hyper-diffusion operators used in (Fourier) $k$-space) allows us to deal with this problem.
Together with a pseudo-spectral representation of the plane perpendicular to the 
background magnetic field, and the option of a spectral-like algorithm for the 
dynamics along the field, the Hermite representation of velocity space implies that \viriato~is ideally suited to the investigation of magnetised kinetic plasma turbulence and magnetic reconnection, with the unique capability of allowing for the direct monitoring of energy flows in phase-space~\cite{loureiro_fast_2013}.

A series of linear and nonlinear numerical tests of \viriato~is presented, 
with emphasis on Orszag-Tang-type decaying turbulence, 
both in the fluid and kinetic limits, where it is shown that \viriato~recovers the 
theoretically expected power-law spectra. In this context,
an interesting, novel result that warrants further investigation and will be discussed in a 
separate publication is the $\sim m^{-1/2}$ velocity-space (Hermite) spectrum that is 
obtained in the 3D kinetic (sub-ion Larmor radius scales) Orszag-Tang run 
presented in~\secref{sec:3D_OT} (see~\fig{fig:OTBW_3D_KAW_gm_spectra}). 
This particular form of the Hermite spectrum is indicative of 
linear phase mixing~\cite{zocco_reduced_2011,loureiro_fast_2013} and suggests that this
(and ensuing Landau damping) may be 
a key energy transfer mechanism in kinetic decaying turbulence.

\section*{Acknowledgements}
The authors are greatly indebted to Alex Schekochihin for many discussions and ideas that have been fundamental to this work.
NFL thanks Paul Dellar for pointing out the high-order Fourier smoothing method of 
Ref.~\cite{hou_computing_2007}, Ravi Samtaney for discussions on high-order 
integration schemes for advection-type partial differential equations, and Ryusuke Numata for providing the data obtained with {\tt AstroGK} that appears in \fig{fig:tear_bench_linear} of this paper.
This work was partly supported by 
Funda\c{c}\~ao para a Ci\^{e}ncia e Tecnologia via 
Grants UID/FIS/50010/2013, PTDC/FIS/118187/2010 and IF/00530/2013,
and by the Leverhulme Trust 
Network for Magnetised Plasma Turbulence.
Simulations were carried out at HPC-FF (Juelich), Helios (IFERC), Edison and Hopper (NERSC),
Kraken (NCSA) and Stampede (TACC).

\section*{Appendix A: Addition of a background electron temperature gradient}
\label{app:gradTe}
A recent paper by Zocco \etal~\cite{zocco_kinetic_2015} extends the KREHM model to include a 
background electron temperature gradient. This extension is also implemented in \viriato; results of 
ongoing investigations exploring different instabilities introduced by these terms (namely, the electron 
temperature gradient mode, and the microtearing instability) will be reported elsewhere. For completeness, 
we write below the KREHM equations with this extension in normalised form (see \secref{sec:KREHM_norms} for 
the details of the normalisation adopted in \viriato). They are:
\bea
&&\frac{d n_e}{d t} =  \[\Apar, \lapperp\Apar\] -\frac{\d}{\d z}\lapperp\Apar,\\
&&\frac{d}{d t}\(\Apar - d_e^2\lapperp\Apar\)  
= \eta\lapperp\Apar + 
\rho_s^2\[n_e+\sqrt{2}g_2,\Apar\] -
\frac{1}{\sqrt{2}}\frac{\rhos}{d_e}\alpha_{Te}\frac{\d \Apar}{\d y}\nonumber\\
&&\qquad\qquad -\frac{\d \varphi}{\d z} + \rhos^2\frac{\d}{\d z}\(n_e + \sqrt{2} g_2\)\\
&&\frac{d g_2}{d t} =
\sqrt{3}\frac{\rho_s}{d_e}\left\{ \[\Apar,g_3\]-\frac{\d g_3}{\d z}\right\} 
+\sqrt{2} \left\{\[\Apar,\lapperp\Apar\]-\frac{\d}{\d z}\lapperp \Apar\right\} \nonumber\\
&&\qquad\qquad- \frac{1}{2}\frac{1}{\rhos d_e}\alpha_{Te}\frac{\d \varphi}{\d y},\\
&&\frac{d g_m}{d t} = 
\sqrt{m+1}\frac{\rho_s}{d_e} \left\{\[\Apar,g_{m+1}\] -\frac{\d g_{m+1}}{\d z}\right\}
+\sqrt{m}\frac{\rho_s}{d_e} \left\{\[\Apar,g_{m-1}\] -\frac{\d g_{m-1}}{\d z}\right\}\nonumber\\
&&\qquad\qquad- m\nu_{ei}g_m 
+\delta_{m,3}\frac{1}{2}\frac{1}{d_e^2}\alpha_{Te}\frac{\d\Apar}{\d y}, \quad m>2,
\eea
where $\alpha_{Te}=\rhoe/L_{Te}L_{\parallel}/L_\perp$, with $\rhoe$ the electron Larmor radius 
and $L_{Te}$ the electron temperature gradient scale length.

\end{document}